\documentclass[]{jfm}

\usepackage{graphicx}
\usepackage{newtxtext}
\usepackage{newtxmath}
\usepackage{natbib}
\usepackage{hyperref}
\hypersetup{
    colorlinks = true,
    urlcolor   = blue,
    linkcolor  = blue,
    citecolor  = black,
}

\newcommand{\RomanNumeralCaps}[1]
\linenumbers

\usepackage{amsmath}
\usepackage{longtable}
\usepackage{enumitem}
\usepackage{siunitx}
\usepackage{physics}
\usepackage{caption}
\usepackage{graphicx}
\usepackage{bm}
\usepackage{float}
\usepackage[dvipsnames]{xcolor}

\newcommand{\changed}[1]{{\color{black} #1}}
\usepackage{array}
\usepackage{tikz}
\usepackage{xcolor}
\usepackage{pgfplots}
\usepackage{graphicx}  
\usepackage{lscape}
\usepackage{rotating}

\definecolor{tabred}{HTML}{d62728}
\definecolor{tabblue}{HTML}{1f77b4}
\definecolor{taborange}{HTML}{ff7f0e}
\definecolor{tabgreen}{HTML}{2ca02c}
\definecolor{tabpurple}{HTML}{9467bd}
\definecolor{lightgreen}{HTML}{67b99a}
\definecolor{darkgreen}{HTML}{036666}
\definecolor{tabolive}{HTML}{bcbd22}

\newcommand{\purplediamond}{\tikz\draw[tabpurple, thick] (0,0.15) -- (0.15,0) -- (0,-0.15) -- (-0.15,0) -- cycle;} 
\newcommand{\blueplus}{\tikz[rotate around={0:(0,0)}]\draw[tabblue, thick] (-0.05,0.15) -- (0.05,0.15) -- (0.05,0.05) -- (0.15,0.05) -- (0.15,-0.05) -- (0.05,-0.05)--(0.05,-0.15) -- (-0.05,-0.15) -- (-0.05,-0.05) -- (-0.15,-0.05) -- (-0.15,0.05) -- (-0.05,0.05) -- cycle;} 
\newcommand{\greenplus}{\tikz[rotate around={0:(0,0)}]\draw[tabgreen, thick] (-0.05,0.15) -- (0.05,0.15) -- (0.05,0.05) -- (0.15,0.05) -- (0.15,-0.05) -- (0.05,-0.05)--(0.05,-0.15) -- (-0.05,-0.15) -- (-0.05,-0.05) -- (-0.15,-0.05) -- (-0.15,0.05) -- (-0.05,0.05) -- cycle;} 

\newcommand{\orangecross}{\tikz[rotate around={45:(0,0)}]\draw[taborange, thick] (-0.05,0.15) -- (0.05,0.15) -- (0.05,0.05) -- (0.15,0.05) -- (0.15,-0.05) -- (0.05,-0.05)--(0.05,-0.15) -- (-0.05,-0.15) -- (-0.05,-0.05) -- (-0.15,-0.05) -- (-0.15,0.05) -- (-0.05,0.05) -- cycle;} 

\newcommand{\bluestar}{%
\tikz\draw[tabblue, thick] (0,0.15) -- (0.05,0.05) -- (0.15,0.05) -- (0.075,-0.025) -- (0.1,-0.15) -- (0,-0.075) -- (-0.1,-0.15) -- (-0.075,-0.025) -- (-0.15,0.05) -- (-0.05,0.05) -- cycle;}

\newcommand{\purpletriangle}{\tikz\draw[tabpurple, thick] (0,0.15) -- (0.15,-0.15) -- (-0.15,-0.15) -- cycle;} 

\newcommand{\bluerottriangle}{\tikz[rotate around={180:(0,0)}]\draw[tabblue,thick] (0,0.15) -- (0.15,-0.15) -- (-0.15,-0.15) -- cycle;} 

\newcommand{\redrottriangle}{\tikz[rotate around={180:(0,0)}]\draw[tabred,thick] (0,0.15) -- (0.15,-0.15) -- (-0.15,-0.15) -- cycle;} 

\usepackage{dashrule}

\definecolor{M030}{HTML}{003554}
\definecolor{M228}{HTML}{006494}
\definecolor{M300}{HTML}{0582ca}
\definecolor{M400}{HTML}{00a6fb}

\title{Variable-property and intrinsic compressibility corrections for turbulence models using near-wall scaling theories}

\author{Asif Manzoor Hasan\aff{1}\corresp{\email{a.m.hasan@tudelft.nl}}, 
Alex Jos\'e Elias\aff{2}, 
Florian Menter\aff{3}
\and Rene Pecnik\aff{1}\corresp{\email{r.pecnik@tudelft.nl}}}
\affiliation{\aff{1}Process \& Energy Department, Delft University of Technology, Leeghwaterstraat 39, 2628~CB, Delft, The Netherlands
\aff{2}ESSS – Engineering Simulation and Scientific Software, Brazil
\aff{3}Ansys Inc., Otterfing 83624, Germany}

\begin{document}
\maketitle

\begin{abstract}
We introduce a novel approach to derive compressibility corrections for Reynolds-averaged Navier-Stokes (RANS) models.
Using this approach, we derive variable-property corrections for wall-bounded flows {that take into account the distinct scaling characteristics of the inner and outer layers, {extending the earlier work of Otero Rodriguez et al. [Int. J. Heat Fluid Flow, 73, 2018].}}
We also propose modifying the eddy viscosity to account for changes in the near-wall damping of turbulence due to intrinsic compressibility effects. {The resulting corrections are consistent with our recently proposed velocity transformation [Hasan et al., Phys. Rev. Fluids, 8, L112601, 2023] in the inner layer and the Van Driest velocity transformation in the outer layer.}
Furthermore, we address some important aspects related to the modeling of the energy equation, primarily focusing on the turbulent Prandtl number and the modeling of the source terms. Compared to the existing state-of-the-art compressibility corrections, the present corrections, combined with accurate modeling of the energy equation, lead to a significant improvement in the results for a wide range of turbulent boundary layers and channel flows. 
The proposed corrections have the potential to enhance modeling across a range of applications, involving low-speed flows with strong heat transfer, fluids at supercritical pressures, and supersonic and hypersonic flows.
\end{abstract}


\section{Introduction}\label{Sec:intro}

Accurate modeling of turbulent compressible and variable-property flows is crucial for a wide range of engineering applications. In particular, turbulence  determines skin friction and heat transfer, and consequently the performance and efficiency of heat exchangers, aerospace vehicles, gas turbines, combustion processes, and high-speed propulsion systems. In order to capture the complex turbulence phenomena associated with these flows, standard turbulence models (developed for incompressible flows) must be modified, which are termed as `compressibility corrections'. 

The earliest compressibility corrections for compressible boundary layers were inspired from the turbulence modeling of shear layers. These corrections focus on incorporating direct intrinsic compressibility terms, such as pressure dilatation and dilatational dissipation, into the turbulent kinetic energy equation. For instance, \cite{zeman1990dilatation} proposed a dilatational dissipation model for high-speed free shear flows, which was later extended to wall-bounded flows by \cite{zeman1993new}. Although dilatational dissipation is typically negligible for wall-bounded flows \citep{huang_coleman_bradshaw_1995,zhang2018direct,sciacovelli2024priori},
applying Zeman's model still improves the results, as it adds just the right amount of extra dissipation to reduce the eddy-viscosity, especially important in cooled-wall boundary layers \citep{rumsey2010compressibility}. 
In other words, the improvement occurs for the wrong reasons and the model does not capture the correct physical mechanisms.
This motivates the need for more physics-based compressibility corrections.   
To achieve this, it is important to first understand and characterize the underlying physics.  

Compressibility effects in wall-bounded turbulent flows can be classified into two main branches. The first involves effects related to heat transfer, often referred to as variable-property effects. The second branch is associated with changes in fluid volume in response to changes in pressure, also termed intrinsic compressibility effects \citep{lele1994compressibility,hasan2024intrinsic}. While variable-property effects can be significant at any, or even at zero, Mach numbers, intrinsic compressibility effects become important only at high Mach numbers. Therefore, it is crucial that compressibility corrections are developed with a clear distinction of these mechanisms and that they remain consistent with the relevant physics—whether concerning heat transfer (variable-property effects) or intrinsic compressibility. 
This can be ensured by deriving separate compressibility corrections from scaling theories associated with these effects.

For flows at non-hypersonic Mach numbers, Morkovin's hypothesis suggests that intrinsic compressibility effects are small, and only mean property (density and viscosity) variations are important to describe the turbulence dynamics \citep{morkovin1962effects}. 
Because of these variations in mean properties, the conventional definitions of friction velocity \changed{($u_\tau = \sqrt{\tau_w/\rho_w}$, with $\tau_w$ the wall shear stress and $\rho_w$ the wall density)} and viscous length scales \changed{($\delta_v=\mu_w/(\rho_w u_\tau)$, with $\mu_w$ the wall viscosity)}, are inaccurate for developing scaling laws. This led to the development of the semi-local scaling framework  \citep{van1951turbulent,morkovin1962effects,huang_coleman_bradshaw_1995, coleman1995numerical, patel2015semi, trettel2016mean, patel2016influence}, where the friction velocity and viscous length scales are defined using local density and viscosity, as 
\begin{equation}\label{Eq.vd,mork}
    \begin{aligned}
 u_\tau^* = \sqrt{\frac{\tau_w}{\bar \rho}}, \,&&&\, \delta_v^* = \frac{\bar \mu}{\bar \rho u_\tau^*}, 
    \end{aligned}
\end{equation}
with $\bar \rho$ and $\bar \mu$ the mean local density and viscosity that vary in the wall-normal direction. 

Several compressible scaling laws in literature are based on these modified scales. For instance, the popular Van Driest velocity transformation accounts for changes in the friction velocity scale as
\begin{equation}\label{vdtrans}
    \bar U_{VD}^+ = \int_0^{\bar u^+} {\sqrt{\frac{\bar \rho}{\rho_w}}} d\bar u^+ = \int_0^{\bar u^+} \frac{u_\tau}{u_\tau^*} d\bar u^+,
\end{equation}
where $\bar u^+ = \bar u/u_\tau$ represents the classically scaled mean velocity.
This transformation, when plotted as a function of $y^+ = y/\delta_v$, leads to a collapse on to the incompressible law of the wall for adiabatic flows, however, its accuracy deteriorates for diabatic flows \citep{bradshaw1977compressible,huang1994van,trettel2016mean,patel2016influence,griffin2021velocity}. This is because the Van Driest transformation does not account for changes in the viscous length scale, which can vary significantly in diabatic flows but remains nearly constant in adiabatic flows close to the wall. 

Subsequently, \cite{trettel2016mean} and \cite{patel2016influence} derived the semi-local velocity transformation, which is an extension to the Van Driest velocity transformation accounting for variations in the semi-local viscous length scale. This transformation (also known as the TL transformation) can be written as
\begin{equation}\label{TL}
    \bar u^* = \bar U_{TL}^+ = \int_0^{\bar u^+} \left(1 - \frac{y}{\delta_v^*} \frac{d\delta_v^*}{dy} \right) {\frac{u_\tau}{u_\tau^*}} d\bar u^+. 
\end{equation}
When plotted as a function of the semi-locally scaled wall-normal coordinate $y^* = y/\delta_v^*$, $\bar u^*$ collapses on to the incompressible law of the wall for low-Mach number \citep{patel2016influence} and moderate-Mach number \citep{trettel2016mean} channel flows. However, this transformation loses accuracy for high-Mach number boundary layers \citep{patel2016influence, trettel2016mean, griffin2021velocity}, where Morkovin's hypothesis fails and intrinsic compressibility effects can no longer be neglected \citep{hasan2023incorporating}. 

At high Mach numbers, intrinsic compressibility effects modify the near-wall damping of turbulent shear stress, leading to an upward shift in the semi-locally transformed mean velocity profile \citep{hasan2023incorporating,hasan2024intrinsic}. 
By adjusting the damping function of a mixing-length turbulence model, \cite{hasan2023incorporating} proposed an extension to the semi-local velocity transformation, given as 
\begin{equation}
    \bar U_{HLPP}^+ = \int_0^{\Bar{u}^+} \! \!
     \left({ \frac{1 + \kappa y^* {D(y^*,M_\tau)}} {1 + \kappa {y^*} {D(y^*,0)}}}\right){\left({1 - \frac{y}{\delta_v^*}\frac{d \delta_v^*}{dy}}\right)}\, {\frac{u_\tau}{u_\tau^*}} \, {d \Bar{u}^+}.
     \label{eq:asif}
\end{equation}
Here, the damping function is given as
\begin{equation}\label{damp}
D(y^*,M_\tau) = \left[1 - \mathrm{exp}\left({\frac{-y^*}{A^+ + f(M_\tau)}}\right)\right]^2,
\end{equation}
with $f(M_\tau) = 19.3 M_\tau$, where $M_\tau = u_\tau/a_w$ is the friction Mach number and $a_w$ is the speed of sound based on wall properties. This transformation, when plotted as a function of $y^*$, collapses on to the incompressible law of the wall for a wide variety of high- and low-speed turbulent flows including (but not limited to) adiabatic and cooled boundary layers, adiabatic and cooled channels, supercritical flows, and flows with non-air-like viscosity laws.

These compressible scaling laws can provide guidelines for developing compressibility corrections for turbulence models. 
For instance, \cite{huang1994turbulence} demonstrated that to achieve the correct slope of the Van Driest scaled mean velocity profile in the logarithmic layer, the model constants must be functions of the mean density gradients. 
Later, \cite{catris2000density} argued that the density dependence from the model constants can be eliminated by modifying the turbulent diffusion term in the turbulence model equations. Consequently, they modified the diffusion terms of several turbulence models, and found that the correct slope of $1/\kappa$ in the Van Driest transformed mean velocity profile was obtained.                                  

A formal approach to deriving compressibility corrections for turbulence models from scaling laws was provided by \cite{pecnik2017scaling}. 
They first scaled the mean momentum and continuity equations using $u_\tau^*$ as the velocity scale and channel half-height $h$ (or $\delta_v$ if one considers the equations in their inner-scaled form) as the length scale. From these, they derived a semi-locally scaled turbulence kinetic energy (TKE) equation and, by analogy, formulated a corresponding semi-locally scaled dissipation equation. 
By rewriting these equations in the dimensional form, \cite{otero2018turbulence} analytically derived variable-property corrections for several turbulence models and noted that these corrections only modify the diffusion terms. 
Interestingly, this modification closely resembles the compressibility corrections proposed by \cite{catris2000density}. The key difference is that in \cite{otero2018turbulence}, the correction applies to the entire diffusion term (both molecular and turbulent), whereas in \cite{catris2000density}, only the turbulent diffusion is corrected. Additionally, the derivation in \cite{otero2018turbulence} results in a slightly different form for the diffusion of turbulent kinetic energy. Despite these differences, both approaches yield very similar results.

The compressibility corrections of \cite{catris2000density} and \cite{otero2018turbulence} (hereafter abbreviated as `CA/OPDP' based on the last names of the authors) both give results consistent with Van Driest's scaling, since they are essentially based on $u_\tau^*$ and $h$ (channel half-height) as the relevant velocity and length scales. 
However, this makes the CA/OPDP corrections valid only in the outer layer, as they do not account for variations in the viscous length scale $\delta_v^*$ in the inner layer. Despite this limitation, the results obtained with the CA/OPDP corrections are still accurate, even for diabatic flows, where the viscous length scales can vary significantly. This is because the variations in the viscous length scale are taken into account differently. Specifically: in the $k$-$\epsilon$ model by using $y^*$ instead of $y^+$ in the damping functions; 
in the Spalart-Allmaras model through the damping function $f_{v1}$ that uses a semi-locally consistent parameter $\chi = \check \nu/\bar \nu = \check \nu/(u_\tau^* \delta_v^*)$; in the $v^2$-$f$ model with the length scale $L_t$ that switches from $k^{1.5}/\epsilon$ (where $k$ is the TKE and $\epsilon$ is its dissipation rate) to the Kolmogorov length scale $\eta$ in the vicinity of the wall which is proportional to $\delta_v^*$ \citep{patel2016influence}. However, this indirect accounting of the variations in the viscous length scale is not robust, and would fail for turbulence models without damping functions, for instance, the $k$-$\omega$ SST model \citep{menter1993zonal}, as observed in \cite{catris2000density} and \cite{otero2018turbulence}. Thus, compressibility corrections that account for viscous length scale variations directly in the model equations
are needed.

At high Mach numbers, corrections based solely on mean property variations are insufficient, as intrinsic compressibility effects also play an important role. These effects modify the near-wall damping of turbulent shear stress, causing it to shift outwards with increasing Mach number. From a turbulence modeling standpoint, this implies that the eddy-viscosity formulation needs to be augmented with a damping function which accounts for this outward shift in the turbulent shear stress as a function of Mach number. In fact, as mentioned earlier, the velocity transformation of \cite{hasan2023incorporating} is based on such a modification of the damping function for a mixing-length model. Similar modifications are also needed for other turbulence models.

In addition to the compressibility corrections discussed earlier, accurate modeling of the energy equation is essential for estimating thermophysical properties, such as density and viscosity, which are critical for solving the turbulence model equations accurately. There are two key aspects of the energy equation modeling: (1) accurate estimation of the eddy-conductivity (analogue of eddy-viscosity) and (2) accurate modeling of the source terms. For high-speed flows, the eddy-conductivity is often estimated using a constant turbulent Prandtl number ($Pr_t$) of 0.9 \citep{wilcox2006turbulence}. However, this constant-$Pr_t$ assumption has been questioned in various recent papers, especially close to the wall \citep{huang2022direct,griffin2023near, chen2024improved}. The source terms, on the other hand, require modeling of the viscous and turbulent diffusion of the mean and turbulent kinetic energy. Using the direct numerical simulation (DNS) data of turbulent channel flows, \cite{huang2023velocity} argued that the viscous and turbulent diffusion of the TKE is negligible. On the contrary, using the DNS data of turbulent boundary layers, \cite{cheng2024} showed that only the third-order correlation (turbulent diffusion of TKE) is negligible, while the viscous diffusion term remains important.

Therefore, the aim of this work is threefold: 
    (1) to properly account for changes in the viscous length scale in the inner layer directly within the turbulence model equations, thereby enabling more accurate predictions in flows with heat transfer at low as well as high Mach numbers;
    (2) to further enhance the model by incorporating intrinsic compressibility effects, allowing for improved predictions for high Mach number flows; and
    (3) to correctly model the energy equation by including the viscous and turbulent diffusion of TKE, leading to better predictions of the temperature profile for high-speed turbulent boundary layers and channel flows. 
The paper is written with a particular emphasis on the $k$-$\omega$ {SST} model \citep{menter1993zonal}. However, the proposed approach can be extended to other models such as the \cite{spalart1992one} model (see Appendix~\ref{Sec:appSA}), $k$-$\epsilon$ models, and the $v^2$-$f$ model \citep{durbin1991near}.

\section{Variable-property corrections}\label{Sec:derivvp}
\subsection{Inner layer}\label{Sec:derivInner}

From several studies in the past decades \citep{morkovin1962effects, coleman1995numerical, huang_coleman_bradshaw_1995, patel2015semi, trettel2016mean, modesti2016reynolds, patel2017scalar, zhang2018direct}, it has been shown that turbulence quantities, when semi-locally scaled (i.e. using $u_\tau^*$ and $\delta_v^*$ as the relevant velocity and length scales, respectively), collapse well onto their respective incompressible counterparts when plotted as a function of the semi-local coordinate $y^*$. 
Some of these quantities, in their classically and semi-locally scaled form are listed in table~\ref{scaletab}.
\begin{table}
\centering\renewcommand{\arraystretch}{1.5}
\begin{tabular}{m{2.4cm} m{5cm} m{5cm} }
Quantity &  Incompressible (classical) & Compressible (semi-local) 
\\ \hline
Wall distance & $y^+ = {y}/{\delta_v}$ & $y^* = {y}/{\delta_v^*}$
\\
Mean shear & ${d\bar u^+}/{d y^+} = ({\delta_v}/{u_\tau}){d\bar u}/{d y} $ & ${d\bar u^*}/{d y^*} = ({\delta_v^*}/{u_\tau^*}){d\bar u}/{d y}$ 
\\ 
TKE & $k^+ = {k}/{u_\tau^2}$ & $k^* = {k}/{{u_\tau^*}^2}$ 
\\ 
{Turb. diss.} & {$\epsilon^+ = {\epsilon}/({{u_\tau}^3/\delta_v})$} & {$\epsilon^* = {\epsilon}/({{u_\tau^*}^3/\delta_v^*})$} 
\\
Spec. turb. diss. & $\omega^+ = {\omega}/({u_\tau/\delta_v})$ & $\omega^* = {\omega}/({u_\tau^*/\delta_v^*})$ 
\\
Eddy visc. & $\mu_t^+ = {\mu_t}/({\rho_w u_\tau \delta_v}) = {\mu_t}/{\mu_w}$ & $\mu_t^* = {\mu_t}/({\bar\rho u_\tau^* \delta_v^*}) = {\mu_t}/{\bar \mu}$ 
\\
Dyn. visc. & $\mu^+ = {\mu_w}/({\rho_w u_\tau \delta_v}) = {\mu_w}/{\mu_w}{ = 1}$ & $\bar \mu^* = {\bar \mu}/({\bar\rho u_\tau^* \delta_v^*}) = {\bar \mu}/{\bar \mu}{ = 1}$ 
\end{tabular}
\captionof{table}{An example of quantities that are classically and semi-locally scaled in the inner layer. {Note that $\bar u^*$ in $d \bar u^*/dy^*$ represents the semi-locally transformed mean velocity, as defined in~\eqref{TL}.}}
\label{scaletab}
\renewcommand{\arraystretch}{1.0}
\end{table}

Similarly, we argue that if the individual variables collapse when semi-locally scaled, then their model equations written in the semi-locally scaled form must be analogous to those written in the classically scaled form for incompressible flows. Here, we enforce a strict analogy by replacing the individual variables in a classically scaled equation by their semi-locally scaled counterparts.

Let us start by writing the modelled turbulence kinetic energy equation in the inner layer of a canonical incompressible {constant-property} flow. Neglecting advection terms, the equation reduces to a simple balance between production, dissipation and diffusion of turbulent kinetic energy as 
\begin{equation}
    \mu_t \left(\frac{d\bar u}{dy}\right)^2 - \beta^\star\rho_w k \omega + \frac{d}{dy}\left[\left(\mu_w + \sigma_k \mu_t\right) \frac{d k}{dy}  \right] = 0,
\end{equation}
where $\mu_t$ is the eddy viscosity, $\bar u$ the mean velocity, $k$ the turbulence kinetic energy, $\omega$ the specific dissipation rate, $\rho_w$, $\mu_w$ the density and viscosity at the wall,
and $\beta^\star$, $\sigma_k$ are the {SST} model constants.

Rewriting this equation using the non-dimensional form of the variables, as given in table~\ref{scaletab}, leads to
\begin{equation}\label{incomprefeqn}
    {\mu_t^+} \left(\frac{d\bar u^+}{dy^+}\right)^2 - \beta^\star k^+ \omega^+ + \frac{d}{dy^+}\left[\left(1 + \sigma_k \mu_t^+\right) \frac{d k^+}{dy^+}  \right] = 0,
\end{equation}
where the superscript `$+$' denotes the classical wall-based scaling. Now we replace all classically scaled variables with their semi-locally scaled counterparts (refer table~\ref{scaletab}), which gives 
\begin{equation}\label{tkestar}
    {\mu_t^*} \left(\frac{d\bar u^*}{dy^*}\right)^2 - \beta^\star k^* \omega^* + \frac{d}{dy^*}\left[\left(1 + \sigma_k \mu_t^*\right) \frac{d k^*}{dy^*}  \right] = 0,
\end{equation}
where the superscript `$*$' denotes semi-local scaling. Rewriting equation~\eqref{tkestar} using the dimensional form of the variables (see table~\ref{scaletab}), we get
\begin{equation}\label{tkestar2}
    \frac{\mu_t}{\bar \mu} \left(\frac{\delta_v^*}{u_\tau^*}\right)^2\left(\frac{d\bar u}{dy}\right)^2 - \beta^\star \frac{k}{{u_\tau^*}^2} \frac{\omega}{u_\tau^*/\delta_v^*} + \frac{d}{d(y/\delta_v^*)}\left[\left(1 + \sigma_k \frac{\mu_t}{\bar \mu}\right) \frac{d (k/{u_\tau^*}^2)}{d(y/\delta_v^*)}  \right] = 0.
\end{equation}
Using the definitions of $u_\tau^*$ and $\delta_v^*$ from equation~\eqref{Eq.vd,mork}, we get
\begin{equation}\label{tkedim1}
    \frac{\mu_t}{\bar \mu} \frac{\bar\mu^2}{\tau_w^2}\left(\frac{d\bar u}{dy}\right)^2 - \beta^\star \frac{\bar \rho}{\tau_w} k \frac{\bar \mu}{\tau_w}\omega + \frac{d}{d(y\sqrt{\tau_w\bar \rho}/\bar\mu)}\left[\left(1 + \sigma_k \frac{\mu_t}{\bar \mu}\right) \frac{d (\bar \rho k/\tau_w)}{d(y\sqrt{\tau_w\bar \rho}/\bar\mu)}  \right] = 0.
\end{equation}
Dividing the equation by $\bar\mu/\tau_w^2$ gives
\begin{equation}\label{tkedim2}
    {\mu_t}\left(\frac{d\bar u}{dy}\right)^2 - \beta^\star {\bar \rho}k \omega + \frac{1}{\bar \mu}\frac{d}{d(y\sqrt{\bar \rho}/\bar\mu)}\left[\left(\bar \mu + \sigma_k {\mu_t}\right)\frac{1}{\bar \mu} \frac{d (\bar \rho k)}{d(y\sqrt{\bar \rho}/\bar\mu)}  \right] = 0.
\end{equation}
{\changed{Applying the chain rule to the diffusion term}, we get
\begin{equation}\label{tkedim3}
    {\mu_t}\left(\frac{d\bar u}{dy}\right)^2 - \beta^\star {\bar \rho}k \omega + \frac{1}{\bar \mu}\frac{dy}{d(y\sqrt{\bar \rho}/\bar\mu)}\frac{d}{dy}\left[\left(\bar \mu + \sigma_k {\mu_t}\right)\frac{1}{\bar \mu} \frac{dy}{d(y\sqrt{\bar \rho}/\bar\mu)}\frac{d (\bar \rho k)}{dy}  \right] = 0,
\end{equation}
where $d(y\sqrt{\bar \rho}/\bar\mu)/dy$ represents the stretching of the wall-normal coordinate due to variable density and viscosity. Note that $y$ in $y\sqrt{\bar\rho}/\bar\mu$ corresponds to the distance from the closest wall. However, $y$ represents this distance only when the origin of the $y$-axis is located on the wall. To make the equations invariant to the choice of the origin of $y$-axis, i.e., to ensure Galilean invariance, we propose to replace $y$ with $\ell$, where $\ell$ represents the distance from the closest wall.

Finally, introducing a stretching variable $S_y$, defined as
\begin{equation}
    S_y = \left(\frac{d \left(\ell\sqrt{\bar\rho}/\bar\mu\right)}{d y}\right)^{-1}=\left(\frac{\sqrt{\bar\rho}}{\bar\mu} + \ell \,\frac{d \left(\sqrt{\bar\rho}/\bar\mu\right)}{d y}\right)^{-1},
\end{equation}
we get
\begin{equation}\label{tkedim4}
    {\mu_t}\left(\frac{d\bar u}{dy}\right)^2 - \beta^\star {\bar \rho}k \omega + \frac{S_y}{\bar \mu}\frac{d}{dy}\left[\left(\bar \mu + \sigma_k {\mu_t}\right)\frac{S_y}{\bar \mu}\frac{d (\bar \rho k)}{dy}  \right] = 0.
\end{equation}
}
Comparing equation~\eqref{tkedim4} with a standard TKE equation for compressible flows, one can note that the present compressibility corrections only modify the modeling of the diffusion term, just as in the CA/OPDP corrections. 
It is important to note that since we enforce a strict analogy, these corrections modify the modeling of the total diffusion term, not just the turbulent diffusion term.

For the ease of implementation in existing computational fluid dynamic solvers, these compressibility modifications can be reformulated in the form of a source term {($\Phi_k^{\textrm{in}}$; where the superscript `in' represents inner layer)} as follows 
\begin{equation}\label{tkedim5}
     {\mu_t}\left(\frac{d\bar u}{dy}\right)^2 - \beta^\star {\bar \rho}k \omega + \frac{d}{d y}\left[\left(\bar \mu + \sigma_k{\mu_t}\right)\frac{d k}{d y}  \right]  + \Phi^{\textrm{in}}_k= 0,
\end{equation}
where the other terms except $\Phi_k^{\textrm{in}}$ represent the standard TKE equation in the inner layer. $\Phi_k^{\textrm{in}}$ can then be written as
\begin{equation}\label{Phi}
\Phi_k^{\textrm{in}} = \frac{S_y}{\bar \mu}\frac{d}{d y }\left[\left(\bar \mu + \sigma_k {\mu_t}\right)\frac{S_y}{\bar \mu} \frac{d (\bar \rho k)}{dy}  \right] - \frac{d}{d y}\left[\left(\bar \mu + \sigma_k {\mu_t}\right)\frac{d k}{d y}  \right].
\end{equation}
$\Phi_k^{\textrm{in}} = 0$ for cases with constant mean properties, while  $\Phi_k^{\textrm{in}}$ is consistent with the CA/OPDP corrections for cases {where $\delta_v^*$ remains uniform in the inner layer.}

Repeating the same procedure for the $\omega$ model equation, we obtain
\begin{equation}\label{phi_om}
 \Phi_\omega^{\textrm{in}} =    \frac{\bar\rho S_y}{\bar \mu^2}\frac{d}{d y }\left[\left(\bar \mu + \sigma_\omega {\mu_t}\right)\frac{S_y}{\bar \mu} \frac{d (\bar \mu \omega)}{d y}  \right]     -\frac{d}{d y}\left[\left(\bar \mu + \sigma_\omega \mu_t\right) \frac{d \omega}{d y}  \right].  
\end{equation}
The standard $k$-$\omega$ {SST} model equations along with the source terms defined in equations~\eqref{Phi}~and~\eqref{phi_om} represent the proposed  compressibility corrections in the inner layer.

\begin{figure}
	\centering	\includegraphics[width=\textwidth]{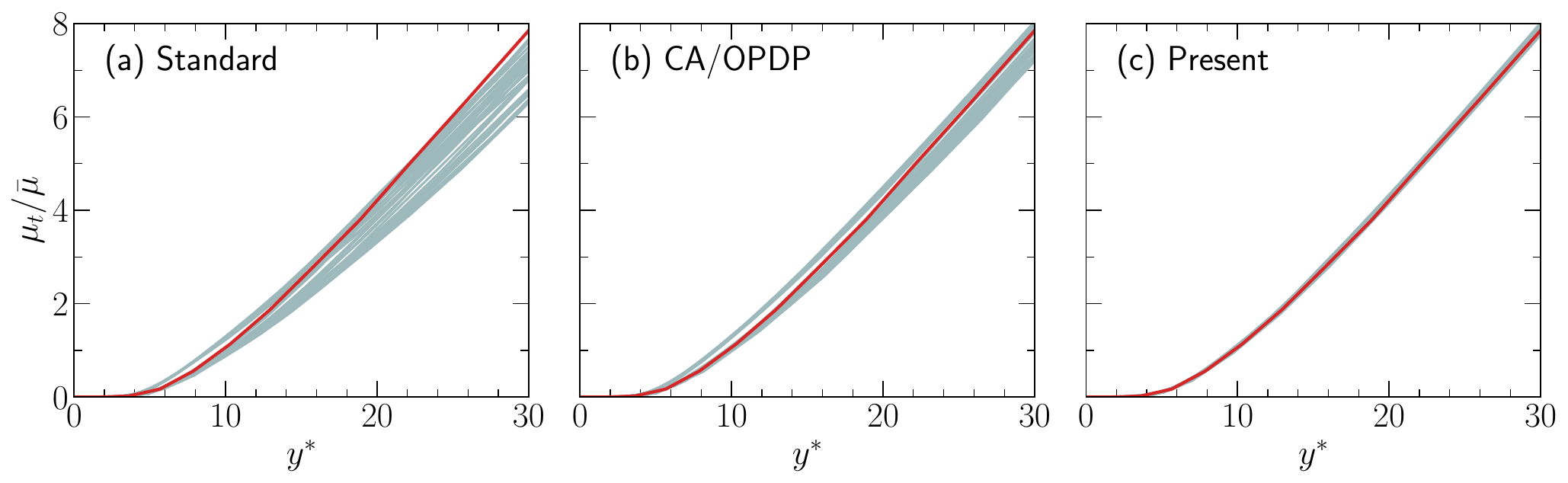}
	\caption{Wall-normal distributions of $\mu_t/\bar \mu$ computed using the $k$-$\omega$ SST model with (a) no corrections, (b) CA/OPDP corrections, and (c) present corrections for the zero-pressure-gradient turbulent boundary layers described in section~\ref{results} (table~\ref{tab:casescfd}).  
 The red lines represent the constant-property case of \cite{sillero2013one} at $Re_\tau=1437$. For details about the implementation, refer section~\ref{implementation}.}
 \label{Fig:mutmu}
\end{figure}

Figure~\ref{Fig:mutmu} shows $\mu_t/\bar\mu$ obtained with the $k$-$\omega$ SST model with (a) no corrections, with (b) CA/OPDP corrections, and with (c) the present corrections for numerous compressible turbulent boundary layers (represented by gray lines) from the literature (see section~\ref{implementation} for details on the implementation and section~\ref{results}, table~\ref{tab:casescfd} for details on the cases). The red lines in the figure show an incompressible case of \cite{sillero2013one} at $Re_\tau = 1437$. The collapse on to a single curve in figure~\ref{Fig:mutmu}(c) clearly shows that the present corrections are consistent with the semi-local scaling framework. 

\subsection{Outer layer}
In the outer layer, the appropriate length scale is $\delta$ (boundary layer thickness; equivalent to $h$ for channels) instead of $\delta_v^*$. Thus, the corrections discussed so far, which are based on $\delta_v^*$, are not valid there.
To obtain valid compressibility corrections in the outer layer, one may follow the same approach as in the previous section, but using $u_\tau^*$ and $\delta$ as the relevant scales \citep{smits2006turbulent,hasan2024estimating}. This leads to {(see Appendix~\ref{Sec:appouter} for a detailed derivation)}
\begin{equation}\label{phikout}
    \Phi_k^{\textrm{out}}  =  \frac{1}{\sqrt{\bar \rho}}\frac{\partial}{\partial y}\left[\left(\bar \mu+\sigma_{k} \mu_t \right)\frac{1}{\sqrt{\bar \rho}} \frac{\partial (\bar \rho k)}{\partial y}  \right] - \frac{\partial}{\partial y}\left[\left(\bar \mu+\sigma_{k} \mu_t\right) \frac{\partial k}{\partial y}\right],
\end{equation}
and
\begin{equation}\label{phiomout}
\Phi_\omega^{\textrm{out}}  =  \frac{\partial}{\partial y}\left[\left(\bar \mu+\sigma_{\omega} \mu_t \right)\frac{1}{\sqrt{\bar \rho}} \frac{\partial (\sqrt{\bar \rho} \omega)}{\partial y}  \right] - \frac{\partial}{\partial y}\left[\left(\bar \mu+\sigma_{\omega} \mu_t\right) \frac{\partial \omega}{\partial y}\right].
\end{equation}
{Note that} these equations are identical to the corrections proposed in \cite{otero2018turbulence}. 

\changed{In the SST model, the transport equation for $\omega$ includes a cross-diffusion term, given by  
\begin{equation}\label{CDkom}
C D_{k \omega}= 2 (1-F_1) \frac{\bar \rho \sigma_{\omega 2}}{\omega} \frac{\partial k}{\partial y} \frac{\partial \omega}{\partial y}, 
\end{equation}
where $F_1$ is a blending function introduced by \cite{menter1993zonal} to smoothly transition between the $k$-$\omega$ and $k$-$\epsilon$ models. This term is primarily active in the outer region of a boundary layer, where $F_1 \to 0$.  
We argue that because the cross-diffusion term originates from the turbulent diffusion term\textemdash and given that our proposed corrections target the diffusion term\textemdash it is necessary to modify this term for consistency.  

Following the same approach as described earlier, the variable-property-corrected cross-diffusion term is given by  
\begin{equation}
C D_{k \omega}= 2 (1-F_1) \frac{\sigma_{\omega 2}}{\sqrt{\bar \rho} \omega} \frac{\partial \bar \rho k}{\partial y} \frac{\partial \sqrt{\bar \rho }\omega}{\partial y}.
\end{equation}  

Finally, as previously done, we express the corrections on the $CD$ term in the form of a source term as
\begin{equation}\label{phicd}
\Phi_{CD} = \underbrace {2 (1-F_1) \frac{\sigma_{\omega 2}}{\sqrt{\bar \rho} \omega} \frac{\partial \bar \rho k}{\partial y} \frac{\partial \sqrt{\bar \rho }\omega}{\partial y}}_{\textrm{variable-property-corrected}} - \underbrace{2 (1-F_1) \frac{\bar \rho \sigma_{\omega 2}}{\omega} \frac{\partial k}{\partial y} \frac{\partial \omega}{\partial y}}_{\textrm{conventional}}. 
\end{equation}  
}

\section{{ Proposed variable-property corrections for the entire boundary layer}}\label{Sec:appblend}
Implementing different corrections for the inner and outer layers requires switching between the two forms or blending them with suitable functions. This is essential to preserve the distinct scaling characteristics of these two regions, as demonstrated in \cite{hasan2024estimating} using mixing-length models. 

\changed{To obtain corrections valid in the entire boundary layer, we propose to blend the inner- (equations~\ref{Phi}~\&~\ref{phi_om}) and outer-layer (equations~\ref{phikout}~\&~\ref{phiomout}) source terms using a suitable blending function $\mathcal{F}$ as
\begin{align}\label{blend1}
      \Phi_k & = \mathcal{F} \Phi_k^{\textrm {in}} +  \left(1-\mathcal{F}\right)\Phi_k^{\textrm{out}} , \\  \label{blend2}
      \Phi_\omega &= \mathcal{F}\Phi_\omega^{\textrm {in}} + \left(1-\mathcal{F}\right)\Phi_\omega^{\textrm{out}},  
\end{align}
where $\mathcal{F}$ transitions from unity in the viscous sublayer to zero in the outer layer. In this paper, we use $F_1$ as the appropriate blending function; however, other options could be more suitable, whose discussion is deferred to a future study.}  

\changed{With these corrections, the $k$ and $\omega$ transport equations in an SST model are modified as:
\begin{align}\label{fullsstk}
    \bar \rho \frac{D k}{Dt} & = \textrm{Conventional TKE terms}+ \Phi_{k}, \\
    \bar \rho \frac{D \omega}{Dt} & = \textrm{Conventional $\omega$ terms} + \Phi_{\omega}+ \Phi_{CD}.\label{fullsstom}
\end{align}
}

{Lastly, three important points are worth noting. First, only the wall-normal component of the diffusion term in equations~\eqref{Phi}~-~\eqref{phiomout} is modified; the wall-parallel components remain unchanged. \changed{This is because semi-local scaling, and thus the corrections derived from it, are concerned with variations only along the wall-normal direction. However, applying these corrections also in the wall-parallel directions should not influence the results, since mean property gradients along those directions are comparatively negligible. The formulation of the corrections in cases where the wall-normal direction does not align with the coordinate axis (for e.g. flow over an inclined wall) is discussed in Appendix~\ref{app:normal}.}
Second, as expected, the proposed corrections vanish in the free-stream region, where the mean properties remain constant. 
Third, in free-shear flows, these corrections\textemdash especially the outer-layer ones\textemdash would have minimal impact, leading to little improvement over an uncorrected model \citep{aupoix2004modelling}. The inner-layer corrections are irrelevant in such flows due to negligible viscous effects.

\section{Intrinsic compressibility corrections}

The semi-local scaling approach presented in the previous section accounts for effects associated with mean density and viscosity variations. However, at higher Mach numbers, in addition to variable property effects, intrinsic compressibility effects also play an important role. These compressibility effects modify the near-wall damping of turbulence, {particularly in the buffer layer}, resulting in an outward shift in the eddy-viscosity profile, and an upward shift in the logarithmic portion of the semi-locally (or TL) transformed mean velocity profile \citep{hasan2023incorporating, hasan2024intrinsic}.

To account for this outward shift in the eddy viscosity, \cite{hasan2023incorporating} modified the Van Driest damping function in a mixing length model (equation~\eqref{damp}) as
\begin{equation}
    \mu_t = {\bar\rho u_\tau^*\kappa yD(y^*,0)} \underbrace{\frac{D(y^*,M_\tau)}{D(y^*,0)}}_{D^{ic}},
\end{equation}
where $\bar \rho u_\tau^* \kappa y D(y^*,0)$ is the semi-local eddy viscosity that accounts for mean property variations alone, and 
$D^{ic}$ is the change in damping caused due to intrinsic compressibility effects. {Note that $D^{ic}$ is mainly active in the buffer layer and approaches unity in the log-layer and beyond.}

Similarly, we propose for the SST model, to multiply the eddy viscosity 
with a damping function that captures the outward shift due to intrinsic compressibility effects as 
\begin{equation}\label{sstic}
     \mu_t = {\underbrace{{\frac{a_1 \bar \rho k}{max(a_1\omega,F_2\Omega)}}}_{\textrm{conventional $\mu_t$}}} ({D^{ic}})_{\textrm{sst}},  
\end{equation}
{where the constant $a_1=0.31$, and $\Omega = d\bar u/dy$ for canonical flows \citep{menter1993zonal}}.
The damping function in this equation is defined as 
\begin{equation}\label{Dic_kom}
 (D^{ic})_{\textrm{sst}} = \frac{D(R_t,M_t)}{D(R_t, 0)}, 
\end{equation}
with
\begin{equation}\label{dampkom}
D(R_t,M_t) =  \left[1 - \mathrm{exp}\left(\frac{-R_t}{K + f(M_t)}\right)\right]^2,
\end{equation}
where $M_t = \sqrt{ 2 k}/\bar a$ ($\bar a$ being the local speed of sound) is the turbulence Mach number, and $R_t = \bar \rho k/(\bar \mu \omega)$ is the turbulence Reynolds number. The constant $K$ controls the region in which the damping function $(D^{ic})_{\textrm{sst}}$ is active (analogous to $A^+$ in equation~\eqref{damp}). To apply the modifications mainly in the buffer layer, we choose $K = 3.5$. 

The function $f(M_t)$ controls how the eddy-viscosity decreases (or shifts outward) with increasing Mach number, analogous to the function $f(M_\tau)$ in equation~\eqref{damp}. 
Based on several high-Mach number DNS from literature we find that for the $k$-$\omega$ {SST} model 
\begin{equation}\label{fmt}
f(M_t) = 0.39 M_t^ {0.77}   
\end{equation}
produces accurate results. Refer to Appendix~\ref{Sec:appfmt} for details on the formulation and tuning of this function.

\section{Modeling the energy equation}\label{sec:energyeq}

We now turn our attention to accurately modeling the energy equation, which provides the density, viscosity, and other thermo-physical properties essential for solving the turbulence model equations discussed in the previous sections. \changed{We mainly focus on the inner layer, as in this region the various terms in the energy equation are large in magnitude, which are likely responsible for the high errors observed in the temperature predictions.} 

The total energy equation in the inner layer of canonical wall-bounded flows can be written as   
\begin{equation}\label{energy1}
\frac{d}{dy} \left(-\bar{q}_y - \bar{\rho} \widetilde{v^{\prime \prime}h^{\prime \prime}}\right)+\frac{d}{dy} \left(\tilde{u} \bar{\tau}_{xy} -\tilde{u} \bar{\rho} \widetilde{u^{\prime \prime} v^{\prime \prime}}\right) +\frac{d}{dy}\left(\overline{u_i^{\prime \prime} \tau^{\prime}_{i j}} - \frac{1}{2} \bar{\rho} \widetilde{v^{\prime \prime} u_i^{\prime \prime} u_i^{\prime \prime}}\right) {- \frac{d \bar p}{dx}} \tilde{u} = 0,
\end{equation}
where $q_y$ is the molecular heat flux in the wall-normal direction, $v$ the wall normal velocity, $h$ the enthalpy, and $\tau_{ij}$ the viscous shear stress.
The first term on the left hand side represents molecular and turbulent diffusion of enthalpy, the second term represents molecular and turbulent diffusion of mean kinetic energy, and the third term represents molecular and turbulent diffusion of turbulent kinetic energy. {The last term represents the work done by the mean pressure gradient in the streamwise direction.
For fully developed channel flows, $d \bar p/dx =-{\tau_w}/{h}$, and for zero-pressure-gradient boundary layers, $d \bar p/dx=0$.} 

We can simplify the second term on the left hand side as
\begin{equation}\label{energy2}
    \frac{d}{dy} \left(\tilde{u} \bar{\tau}_{xy} -\tilde{u} \bar{\rho} \widetilde{u^{\prime \prime} v^{\prime \prime}}\right)  = \tau_{tot}\frac{d \tilde{u}}{dy} + \tilde{u}\frac{d \tau_{tot}}{dy},
\end{equation}
where $\tau_{tot} = \bar{\tau}_{xy} - \bar{\rho} \widetilde{u^{\prime \prime} v^{\prime \prime}}$.
For fully developed channel flows, we can write ${d \tau_{tot}}/{dy} = -{\tau_w}/{h} = {d\bar p/dx}$, and for ZPG boundary layers ${d \tau_{tot}}/{dy} = 0$ {in the inner layer}. Taking these simplifications into account, we get 
\begin{equation}\label{energy3}
 \frac{d}{dy} \left(-\bar{q}_y - \bar{\rho} \widetilde{v^{\prime \prime}h^{\prime \prime}}\right)+ (\bar{\tau}_{xy} - \bar{\rho} \widetilde{u^{\prime \prime} v^{\prime \prime}}) \frac{d \tilde{u}}{dy} +\frac{d}{dy}\left(\overline{u_i^{\prime \prime} \tau^{\prime}_{i j}} - \frac{1}{2} \bar{\rho} \widetilde{v^{\prime \prime} u_i^{\prime \prime} u_i^{\prime \prime}}\right) = 0.
\end{equation}
The terms in equation~\eqref{energy3} are modeled as follows \citep{wilcox2006turbulence}:
\begin{equation}\label{energy4}
-\frac{d\bar{q}_y}{dy}  = \frac{d}{dy}\left(\frac{\bar \mu c_p}{Pr} \frac{d \bar T}{dy}\right), \quad
-\frac{d\bar{\rho} \widetilde{v^{\prime \prime}h^{\prime \prime}}}{dy}  =  \frac{d}{dy}\left(\frac{ \mu_t c_p}{Pr_t} \frac{d \bar T}{dy}\right),\\ 
\end{equation}

\begin{equation}\label{energy4b}
\tau_{xy}\frac{d \tilde{u}}{dy}  = \bar \mu \left(\frac{d \tilde{u}}{dy}\right)^2, \quad
-\bar{\rho} \widetilde{u^{\prime \prime} v^{\prime \prime}}\frac{d \tilde{u}}{dy} = \mu_t \left(\frac{d \tilde{u}}{dy}\right)^2,\\ 
\end{equation}
where $Pr_t$ is the turbulent Prandtl number.
The last term on the left hand side in equation~\eqref{energy3} represents the total diffusion of TKE, which is modeled as (see section 2.1, equation~\ref{tkedim4})
\begin{equation}\label{energy5}
\frac{d}{d y}\left(\overline{u_i^{\prime \prime} \tau^{\prime}_{i j}} - \frac{1}{2} \bar{\rho} \widetilde{v^{\prime \prime} u_i^{\prime \prime} u_i^{\prime \prime}}\right)= \frac{S_y}{\bar \mu}\frac{d}{d y }\left[\left(\bar \mu + \sigma_k {\mu_t}\right)\frac{S_y}{\bar \mu} \frac{d (\bar \rho k)}{dy}  \right].  
\end{equation}

With these modeled terms, equation~\eqref{energy3} can be written as
\begin{equation}\label{energy7}
\frac{d}{dy}\left(\left[\frac{\bar \mu c_p}{Pr}+\frac{\mu_t c_p}{Pr_t}\right] \frac{d \bar T}{dy}\right) = -\left(\bar \mu + \mu_t \right) \left(\frac{d \tilde{u}}{dy}\right)^2 - \frac{S_y}{\bar \mu}\frac{d}{d y }\left[\left(\bar \mu + \sigma_k {\mu_t}\right)\frac{S_y}{\bar \mu} \frac{d (\bar \rho k)}{dy}  \right].
\end{equation}
{In the inner layer, the total diffusion of TKE (last term in equation~\ref{energy7}) is balanced by its production and dissipation (see equation~\ref{tkedim4}), such that
\begin{equation}\label{energy7.5}
    \frac{S_y}{\bar \mu}\frac{d}{dy}\left[\left(\bar \mu + \sigma_k {\mu_t}\right)\frac{S_y}{\bar \mu} \frac{d (\bar \rho k)}{dy}  \right] = -{\mu_t}\left(\frac{d\bar u}{dy}\right)^2 + {\bar \rho} \epsilon.
\end{equation}
Using this relation to substitute the total diffusion of TKE in~\eqref{energy7}, we get
\begin{equation}\label{energy8}
\frac{d}{dy}\left(\left[\frac{\bar \mu c_p}{Pr}+\frac{\mu_t c_p}{Pr_t}\right] \frac{d \bar T}{dy}\right) = -\left(\bar \mu + \mu_t \right) \left(\frac{d \tilde{u}}{dy}\right)^2 + \mu_t \left(\frac{d \tilde{u}}{dy}\right)^2 - \bar \rho \epsilon ,
\end{equation}
which further simplifies to
\begin{equation}\label{energy9}
\frac{d}{dy}\left(\left[\frac{\bar \mu c_p}{Pr}+\frac{\mu_t c_p}{Pr_t}\right] \frac{d \bar T}{dy}\right)=
- \underbrace{\bar \mu \left(\frac{d \tilde{u}}{dy}\right)^2 - \bar \rho \epsilon}_{\Phi_e}, 
\end{equation}
where $\Phi_e$ represents the source term in this equation.

Commonly, the TKE dissipation term in~\eqref{energy9} is assumed to be equal to the production term in the inner layer, thereby implying equilibrium, i.e., $\bar \rho\epsilon \approx \mu_t (d\bar u/dy)^2$ \citep{larsson2016large, bose2018wall,huang2023velocity}.
\changed{In other words, the TKE diffusion term in equation~\eqref{energy7} is assumed to be zero.}
However, this assumption breaks down near the wall and is responsible for inaccuracies in predicting the temperature peak in high-speed, cooled-wall boundary layers (see section~\ref{results}). This implies that accurate modeling of $\epsilon$ \changed{(or the TKE diffusion term) in the inner layer} is essential for accurate temperature predictions.}

Another possible source of error in the energy equation~\eqref{energy9} is the turbulent Prandtl number, which is often assumed to be a constant and equal to 0.9 \citep{wilcox2006turbulence}. However, it is well-known that for cooled-wall boundary layers $Pr_t$ is not a constant but rather varies substantially in the near-wall region. Furthermore, at the location of the temperature peak, $Pr_t$ is undefined \citep{griffin2023near,chen2024improved}. In section~\ref{results}, we will briefly analyse the sensitivity of our results with respect to different values of $Pr_t$.

\section{Implementation}\label{implementation}
The proposed corrections primarily differ from the CA/OPDP corrections within the inner layer, while they remain identical in the outer layer (except for the cross-diffusion term discussed in section~\ref{Sec:appblend}). Consequently, for a meaningful comparison between the two approaches, it suffices to focus on solving the inner layer.
\changed{This also allows us to gauge the effectiveness of the proposed inner-layer corrections without any potential influence from the outer layer.}

\changed{In this section, we present the implementation of the proposed corrections for} simple canonical flows that can be modeled as one-dimensional problems, such as the inner layer of zero-pressure-gradient (ZPG) boundary layers and fully developed channel flows. 
\changed{In these flows, mainly the effect of $\Phi_k^{\text{in}}$, $\Phi_\omega^{\text{in}}$, and the damping function $D^{ic}_{\text{sst}}$ is tested.} 
\changed{Subsequently, in section~\ref{full2d}, we test the full corrections, i.e., $\Phi_k^{\text{in}}$ and $\Phi_\omega^{\text{in}}$ in the inner layer, blended with $\Phi_k^{\text{out}}$ and $\Phi_\omega^{\text{out}}$ in the outer layer (equations~\ref{blend1}~and~\ref{blend2}), together with the damping function $D^{ic}_{\text{sst}}$ and the correction for the cross-diffusion term $\Phi_{CD}$.}

\subsection{Zero-pressure-gradient boundary layers} \label{impl_bl}
{To solve the inner layer of boundary layers, we follow the methodology outlined in section 4.6.3 of \cite{wilcox2006turbulence} for incompressible flows. This approach involves solving the integrated streamwise momentum equation, also known as the total stress balance equation, in conjunction with one-dimensional turbulence model equations to determine the inner-layer velocity profile in a turbulent boundary layer. In this work, we extend this framework to compressible flows.}

{For compressible boundary layers, the inner-layer velocity and temperature profiles are obtained by solving the integrated forms of the momentum and energy equations, given as
\begin{equation}\label{tbl:eqns}
\begin{aligned}
\frac{d \bar{u}}{d y}=&\frac{\tau_w}{\bar{\mu}+{\mu_t}},  \quad \text{and} \\ c_p\left(\frac{\bar \mu}{Pr}+\frac{\mu_t}{Pr_t}\right) \frac{d \bar T}{dy} =& -\frac{\mu_w c_p}{Pr}\left(\frac{d \bar T}{dy}\right)_w  - \int_0^{y} \Phi_e
\, dy,
\end{aligned}
\end{equation}
respectively. These equations are solved in a domain spanning from $y=0$ to $y=0.2\delta$, where $y = 0.2\delta$ is arbitrarily taken to be the edge of the inner layer \citep{orlu2010near}. 

The dynamic viscosity $\bar \mu$ in~\eqref{tbl:eqns} is computed using Sutherland's law, the mean density is obtained as $\bar \rho / \rho_w = T_w/\bar T$, and the Prandtl number $Pr$ is equal to 0.72.
The first term on the right-hand-side of the energy equation, scaled by wall-based quantities ($\rho_w u_\tau c_p T_w$), corresponds to the non-dimensional wall heat flux $B_q$.}

{The eddy viscosity $\mu_t$ in~\eqref{tbl:eqns} is computed by} 
solving the one-dimensional SST model (without advection terms), analogous to the one-dimensional $k$-$\omega$ model described in \cite{wilcox2006turbulence} (see equation~4.184). This model is solved with the source terms $\Phi_k^{\textrm{in}}$ and $\Phi_\omega^{\textrm{in}}$ described in equations~\eqref{Phi}~and~\eqref{phi_om}, along with the standard eddy viscosity formulation being multiplied by $(D^{ic})_{\textrm{sst}}$~\eqref{Dic_kom}. 
The boundary conditions for the SST model are defined as
\begin{equation}\label{bc_kom}
\begin{array}{ccccc}
k     = 0,  & \omega = \dfrac{60 \bar \mu}{\beta_1 \bar \rho {(\Delta y)}^2} & \text { at } &  y = 0, & \text{and}\\
k     =  \dfrac{{u_\tau^*}^2}{\sqrt{\beta^\star}}, & \omega =  \dfrac{{u_\tau^*}}{\sqrt{\beta^\star} \kappa y} & \text { at } &  y = 0.2\delta , &
\end{array}
\end{equation}
where {$\Delta y$ is the distance to the next grid point away from the wall}.
The first row of eq.~\eqref{bc_kom} corresponds to the wall boundary condition described in \cite{menter1993zonal}, and the second row corresponds to the log-layer asymptotic solution for $k$ and $\omega$ as described in \cite{wilcox2006turbulence} (see equation~4.185), but adapted to compressible flows.

The turbulent Prandtl number $Pr_t$ in~\eqref{tbl:eqns} is assumed to be a constant and equal to 0.9. However, to show the sensitivity of the results with respect to the constant $Pr_t$ assumption, we will also compute some of the results with $Pr_t$ from  DNS and then compare them with $Pr_t=0.9$.

{Lastly, the source term $\Phi_e$ in~\eqref{tbl:eqns} was defined in equation~\eqref{energy9} as 
\begin{equation}
 \Phi_e = \bar \mu (d \bar u/dy)^2 + \bar \rho \epsilon.   
\end{equation} 
As discussed earlier, a common way to model the TKE dissipation rate is to assume it to be equal to the production term ($\epsilon=\mu_t (d \bar u/dy)^2 $), such that the source term becomes
\begin{equation}\label{phie1}
    \Phi_{e,1} = \bar \mu (d \bar u/dy)^2 + \mu_t (d \bar u/dy)^2,
\end{equation}
where `$1$' in the subscript corresponds to the first type of approximation that we will use in this paper. This approximation is inaccurate near the wall, since close to the wall the production term tends to zero, whereas the dissipation term is finite and balances the total diffusion of TKE.

Therefore, we seek a more accurate representation of $\epsilon$. One way is to use the TKE dissipation rate estimated by the model itself, which in the case of SST is equal to $\epsilon_{\textrm{sst}} = \beta^\star k \omega$. However, just like the production term, $\epsilon_{\textrm{sst}}$ also approaches zero at the wall, as $k \to 0$ while $\omega$ remains finite (see wall-boundary conditions in equation~\ref{bc_kom}). Consequently, using $\epsilon$ from the SST model would yield results similar to those obtained with $\Phi_{e,1}$.
To address this inherent limitation, we follow the approach of \cite{rahman2012exploring} and model the dissipation rate as  
\begin{eqnarray}\label{eps_epsw}
    \epsilon_{\textrm{eff}} = \sqrt{\epsilon_{\textrm{sst}}^2 + \epsilon_w^2},
\end{eqnarray}  
where $\epsilon_w$ ensures non-zero dissipation rate at the wall, given by $\epsilon_w = 2 A_\epsilon ({\bar \mu}/{\bar \rho}) ({d \bar u}/{dy})^2$, with $A_\epsilon = 0.09$ \citep{rahman2012exploring}, and where $\epsilon_{\textrm{eff}}$ represents the effective dissipation rate. With this, the second approximation of $\Phi_e$ that we will use in this paper is given as
\begin{equation}\label{phie2}
    \Phi_{e,2} = \bar \mu (d \bar u/dy)^2 + \bar \rho \epsilon_{\textrm{eff}}.
\end{equation}
}

We solve the equations listed above iteratively until convergence; see \cite{jupnotebook} for more details on the solver.

\subsection{Fully developed channel flows}
For channel flows, instead of solving only for the inner layer, we solve for the entire domain {(spanning from $y=0$ to $y=2h$, $h$ being the channel half-height)}. However, we do not switch to the outer-layer compressibility corrections, but consistently use the inner-layer corrections in the entire domain. This is because the velocity profile in the outer layer of channel flows closely follows the logarithmic profile of the inner layer, meaning the error introduced by not solving the outer layer with the correct corrections is minimal. 

{For fully developed channel flows, we solve the momentum and energy equations, given as
\begin{equation}\label{ranschannel}
\begin{aligned}
\frac{d}{d y}\left[\left({\bar \mu}+\mu_t\right) \frac{d \bar u}{d y}\right] & = - \frac{\tau_w}{h}, & \quad&
\frac{d}{dy}\left(\left[\frac{\bar \mu c_p}{Pr}+\frac{\mu_t c_p}{Pr_t}\right] \frac{d \bar T}{dy}\right)= -\Phi_e,
\end{aligned}
\end{equation}
where $\bar \mu$ is computed using a power-law $(\bar T/T_w)^n$ with an exponent of $0.75$ for high-Mach number flows and $0.7$ for the low-Mach number cases. The mean density is computed as $\bar \rho/\rho_w = T_w/\bar T$, and the Prandtl number $Pr$ is assumed to be a constant and equal to $0.72$ for the high-Mach number cases, and is defined as $Pr = (\bar T/T_w)^{0.7}$ for the low-Mach number cases. The different choices for $\bar{\mu}$ and $Pr$ are consistent with those used in the respective DNSs of these cases.

The eddy-viscosity $\mu_t$ in~\eqref{ranschannel} is obtained by solving the one-dimensional SST model with the proposed corrections (similar to ZPG boundary layers discussed above), along with the wall boundary conditions ($k=0$ and $\omega = {60 \bar \mu}/[\beta_1 \bar \rho (\Delta y)^2]$) imposed on the two walls located at $y=0$ and $y=2h$.

The turbulent Prandtl number ($Pr_t$) in equation~\eqref{ranschannel} is assumed to be constant, with 
$Pr_t = 0.9$ for high-Mach-number channel flows, similar to ZPG boundary layers, and $Pr_t = 1$ 
for low-Mach-number flows. The higher value for the low-Mach-number cases is based on the results of 
\cite{patel2017scalar}, where $Pr_t$ is reported to be close to 1, or even slightly higher for these cases. This is likely due to the strong similarity between momentum and energy equations in these flows.

The source term $\Phi_e$ in~\eqref{ranschannel} for high-Mach number channel flows can be approximated as $\Phi_{e,1}$ or $\Phi_{e,2}$, similar to ZPG boundary layers. 
In contrast, for low-Mach-number flows, the dissipation of mean and turbulent kinetic energy does not contribute to the right-hand side of equation~\eqref{ranschannel}.
This can be explained as follows: $\Phi_{e,1}$ (or $\Phi_{e,2}$) scales with $\rho_{r}  \mathcal{U}^3/\mathcal{L}$, where $\rho_{r}$ is a reference density scale, and $\mathcal U$ and $\mathcal{L}$ are relevant velocity and length scales. respectively. On the other hand, the left-hand side of the energy equation~\eqref{ranschannel} scales with $\rho_{r} \mathcal{U} c_{p,r} T_{r}/\mathcal{L}$, where $c_{p,r}$ and $T_{r}$ are reference scales for $c_p$ and $\bar T$, respectively. The ratio of these scales corresponds to an Eckert number, defined as $\mathcal{U}^2/(c_{p,r} T_r)$ which is proportional to the square of the Mach number. For flows at low (or zero) Mach numbers, this ratio tends to zero which explains why $\Phi_{e,1}$ and $\Phi_{e,2}$ are ineffective in equation~\eqref{ranschannel}. Thus, in those flows, temperature and hence property variations are created by adding a user-defined heat source which is uniform in the domain. 
For the two low-Mach-number gas-like cases considered here, this external heat source is equal to $17.55 (\mu_w c_p/ Pr_w) T_w/h$ \citep{patel2015semi} and $75 (\mu_w c_p/ Pr_w) T_w/h$ \citep{pecnik2017scaling}. }

All the equations described above are solved in their wall-scaled form.  
For more details on the solver, refer the jupyter notebook~\citep{jupnotebook}.


\begin{table}
        \centering
        \rotatebox{0}{
        \begin{tabular}{           m{3.8cm} >{\centering\arraybackslash}m{1.8cm} >{\centering\arraybackslash}m{1.5cm} >{\centering\arraybackslash}m{1.8cm} >{\centering\arraybackslash}m{1.5cm} >{\centering\arraybackslash}m{1.0cm}}
    Source & $Re_{\tau}$ & $M_{\infty}\,,\,M_b$& $M_{\tau}$& $T_w/T_r$ & Symbol\\ \hline
    \textbf{Boundary Layer} &&&&&\\
    \cite{bernardini2011wall} & $205-1113$ & $2-4$ & $0.065-0.105$& $1$ &\scalebox{0.9}{\bluestar}   \\
    \cite{zhang2018direct} & $450-646$& $2.5-13.64$& $0.085-0.195$ & $0.18-1$&\scalebox{0.9}{\blueplus}  
	\\
    \cite{ceci2022numerical} & $500-1000$ & $5.84$& $0.149-0.167$& $0.25$ & \scalebox{0.9}{\purplediamond} \\
    \cite{huang2022direct}  &  $775, 1172$ & $10.9$& $0.167, 0.182$& $0.2$&\scalebox{0.9}{\greenplus}\\
    \cite{cogo2022direct} &$453, 1947$ & $2, 5.86$& $0.066-0.132$& $0.76$& \scalebox{0.9}{\orangecross}\\
    \cite{cogo2023assessment} & $443$ & $2-6$& $0.075-0.159$& $0.35-1$& \scalebox{0.9}{\orangecross}\\
    A. Ceci (Private Comm.) &$450, 475$ & $5.84, 7.87$& $0.129, 0.152$& $0.48, 0.76$&\scalebox{0.9}{\purpletriangle} \\

    \textbf{Channel} &&&&&\\
    \cite{trettel2016mean} &$322-1876$ & $0.7-4$& $0.036-0.118$& $-$ &\scalebox{0.9}{\bluerottriangle}\\
    \cite{patel2016influence} &$395$ & $0$& $0$& $-$ &\scalebox{0.9}{\redrottriangle}\\
    \cite{pecnik2017scaling} & $950$ & $0$& $0$& $-$&\scalebox{0.9}{\redrottriangle}\\
        \end{tabular}}
        \captionof{table}{{Description of the 39 boundary layer and 11 channel flow cases presented in this paper. $Re_\tau = \rho_w u_\tau \delta/\mu_w$ is the friction Reynolds number based on the boundary layer thickness ($\delta$) or the channel half-height ($h$). $M_{\infty} = U_\infty / \sqrt{\gamma R T_\infty}$ is the free-stream Mach number (for boundary layers), $M_{b} = U_b / \sqrt{\gamma R T_w}$ is the bulk Mach number (for channels) and $M_{\tau} = u_\tau / \sqrt{\gamma R T_w}$ is the wall friction Mach number. $T_w/T_r$ is the wall-cooling parameter, with $T_r$ being the adiabatic temperature. These cases are visually represented in figure~\ref{Fig:casescfd}.}}
        \label{tab:casescfd}
\end{table}

\begin{figure}
	\centering	\includegraphics[width=\textwidth]{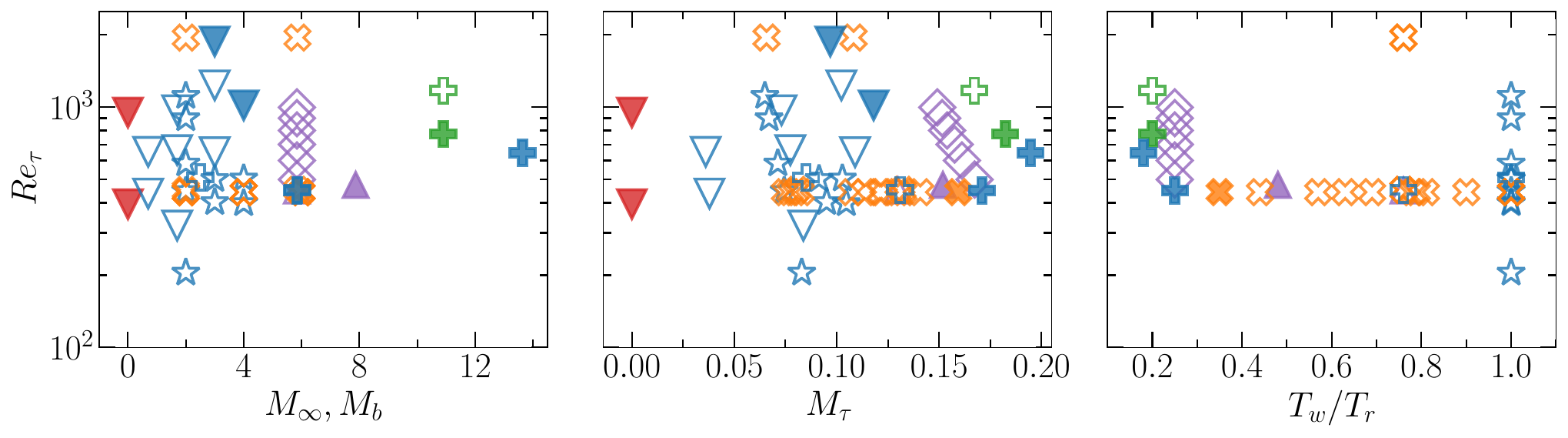}
	\caption{{Friction Reynolds number ($Re_\tau$), free-stream ($M_\infty$) or bulk Mach number ($M_{b}$), friction Mach number ($M_\tau$), and wall-cooling ratio ($T_w/T_r$; where $T_r$ is the adiabatic temperature) for 39 turbulent boundary layers and 11 channel flows described in table~\ref{tab:casescfd}. Filled symbols correspond to the cases whose velocity and temperature profiles are plotted in figures~\ref{Fig:tblcfd}~and~\ref{Fig:chancfd}. Note that $T_w/T_r$ is reported only for boundary layers.}}
 \label{Fig:casescfd}
\end{figure}

\section{{Results}}\label{results}
We now present the results obtained from our inner-layer compressibility corrections~(equations~\ref{Phi},~\ref{phi_om}~and~\ref{Dic_kom}) and compare them with those obtained using the state-of-the-art CA/OPDP corrections \citep{catris2000density,pecnik2017scaling,otero2018turbulence}, which correspond to the outer-layer corrections presented in equations~\eqref{phikout}~and~\eqref{phiomout}. We also present results computed using the baseline model without any corrections.

These results are presented for 39 ZPG boundary layers and 11 channel flows described in table~\ref{tab:casescfd}, and visually represented in figure~\ref{Fig:casescfd}. \changed{Note that for ZPG boundary layers, the results are presented only in the inner layer. For results covering the entire boundary layer, refer to section~\ref{full2d}.}

\subsection{{Zero-pressure-gradient turbulent boundary layers}}

Out of the 39 cases, five cases with increasing wall cooling are selected and their velocity and temperature profiles are shown in figure~\ref{Fig:tblcfd}(a)~and~(b), respectively. 
The different line types correspond to the results obtained with different compressibility corrections and with different modeling approximations in the energy equation, as shown in the legend. They are summarized as follows.
(1) The grey dotted lines correspond to the results obtained with the baseline SST model (without any compressibility corrections), along with $\Phi_{e,1}$ in the energy equation. 
(2) The grey short-dashed lines represent the results computed with the state-of-the-art CA/OPDP compressibility corrections, with $\Phi_{e,1}$ in the energy equation.  
(3) The grey long-dashed lines signify the results when the inner-layer corrections described in equations~\eqref{Phi},~\eqref{phi_om}~and~\eqref{Dic_kom} are used, along with $\Phi_{e,1}$ in the energy equation. 
(4) Finally, the solid grey lines depict the results obtained using the present inner-layer corrections, but with $\Phi_{e,2}$ in the energy equation. All of these results are computed with $Pr_t=0.9$.

\begin{figure}
	\centering	\includegraphics[width=\textwidth]{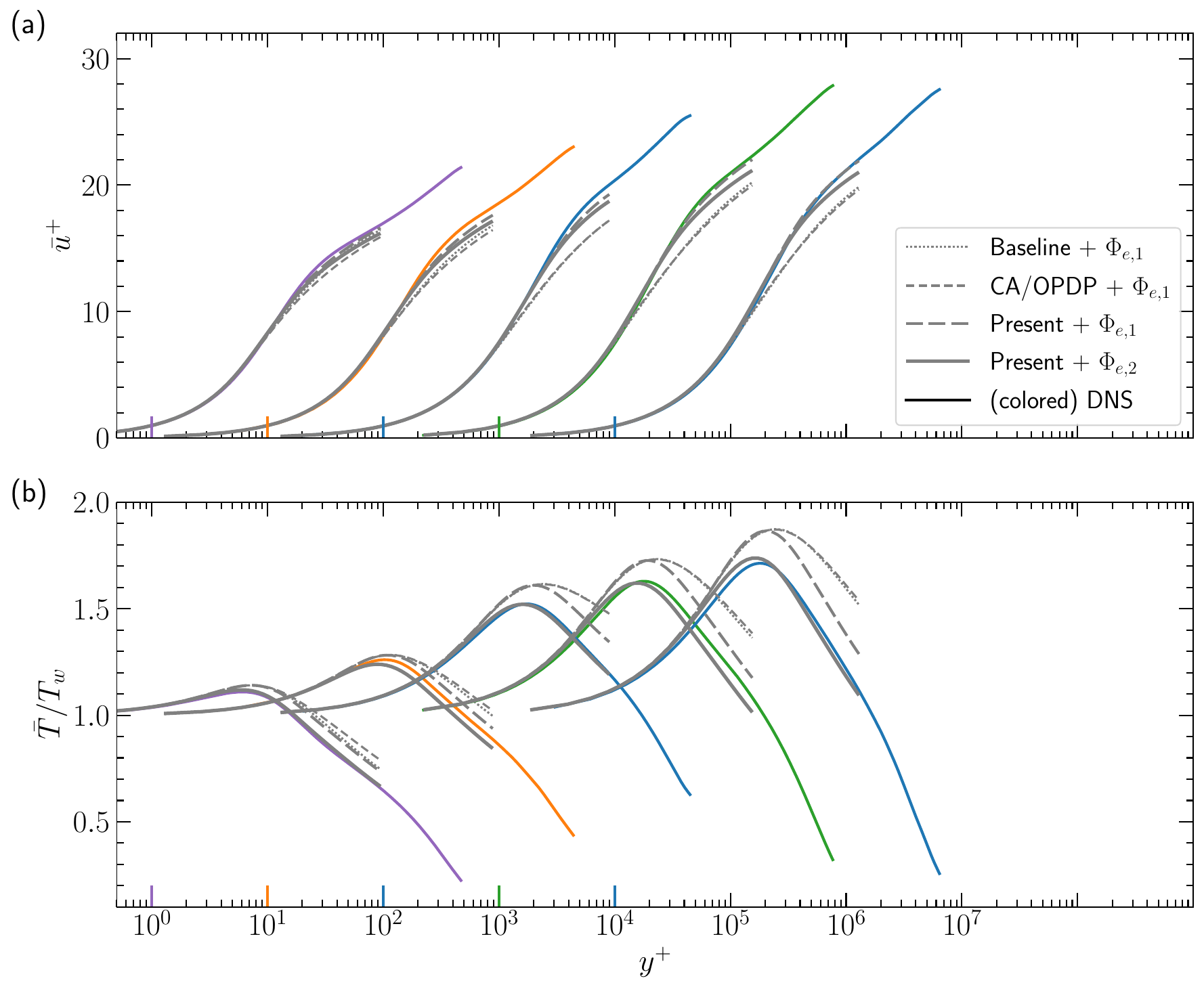}
\caption{{Computed mean velocity (a) and temperature (b) profiles compared to the DNS ({colored} solid lines) for the following boundary layers with increasing wall-cooling: (left to right) $M_\infty=7.87$, $T_w/T_r=0.48$ (A.~Ceci, private communication); $M_\infty=6$, $T_w/T_r=0.35$ \citep{cogo2023assessment}; $M_\infty=5.84$, $T_w/T_r=0.25$ \citep{zhang2018direct}; $M_\infty=10.9$, $T_w/T_r=0.2$ \citep{huang2022direct}; $M_\infty=13.64$, $T_w/T_r=0.18$ \citep{zhang2018direct}. The line colors match the color of the symbols for the respective authors reported in table~\ref{tab:casescfd}.
Refer to the legend for line types. `Baseline' stands for the SST model without corrections, `CA/OPDP' stands for the compressibility corrections proposed in \cite{catris2000density, pecnik2017scaling, otero2018turbulence}, `Present' stands for the corrections proposed in this paper (equations~\ref{Phi},~\ref{phi_om}~and~\ref{Dic_kom}). For clarity, the velocity and temperature profiles for different cases are shifted by one decade along the abscissa. The colored vertical lines on the abscissa signify $y^+=10^0$ for each case, with their colors matching the corresponding cases. 
  }}
 \label{Fig:tblcfd}
\end{figure}

Let us first focus on the grey dotted lines. With increasing wall cooling (see cases from left to right), the velocity profiles computed using the baseline model shift downwards relative to the DNS (solid colored lines), while the temperature profiles are over-predicted.

When the CA/OPDP corrections are applied (see short-dashed lines), the results remain largely similar to the baseline model (also observed previously in \citet{catris2000density, rumsey2010compressibility} and 
\cite{otero2018turbulence}), with slight deterioration observed in some cases.
To further improve the accuracy, it becomes important to account for the variations in the viscous length scale ($\delta_v^*$) in the inner layer. Moreover, the cases in figure~\ref{Fig:tblcfd} also correspond to higher Mach numbers, and thus, it also becomes important to account for intrinsic compressibility effects. 

Upon applying the present corrections (see long-dashed lines), which account for $\delta_v^*$ variations and intrinsic compressibility effects, the results for both velocity and temperature substantially improve with respect to the baseline and CA/OPDP corrections. However, the temperature profiles are still inaccurate compared to the DNS, mainly for the strongly-cooled cases.

This inaccuracy in the temperature profiles can be resolved by replacing $\Phi_{e,1}$ in the energy equation with $\Phi_{e,2}$ (grey solid lines). Clearly, the temperature profiles with $\Phi_{e,2}$ are more accurate, particularly in capturing the peak temperature. This highlights the importance of accurately estimating the source terms in the energy equation for accurate temperature estimations in high-speed boundary layers.

\begin{figure}
	\centering	\includegraphics[width=\textwidth]{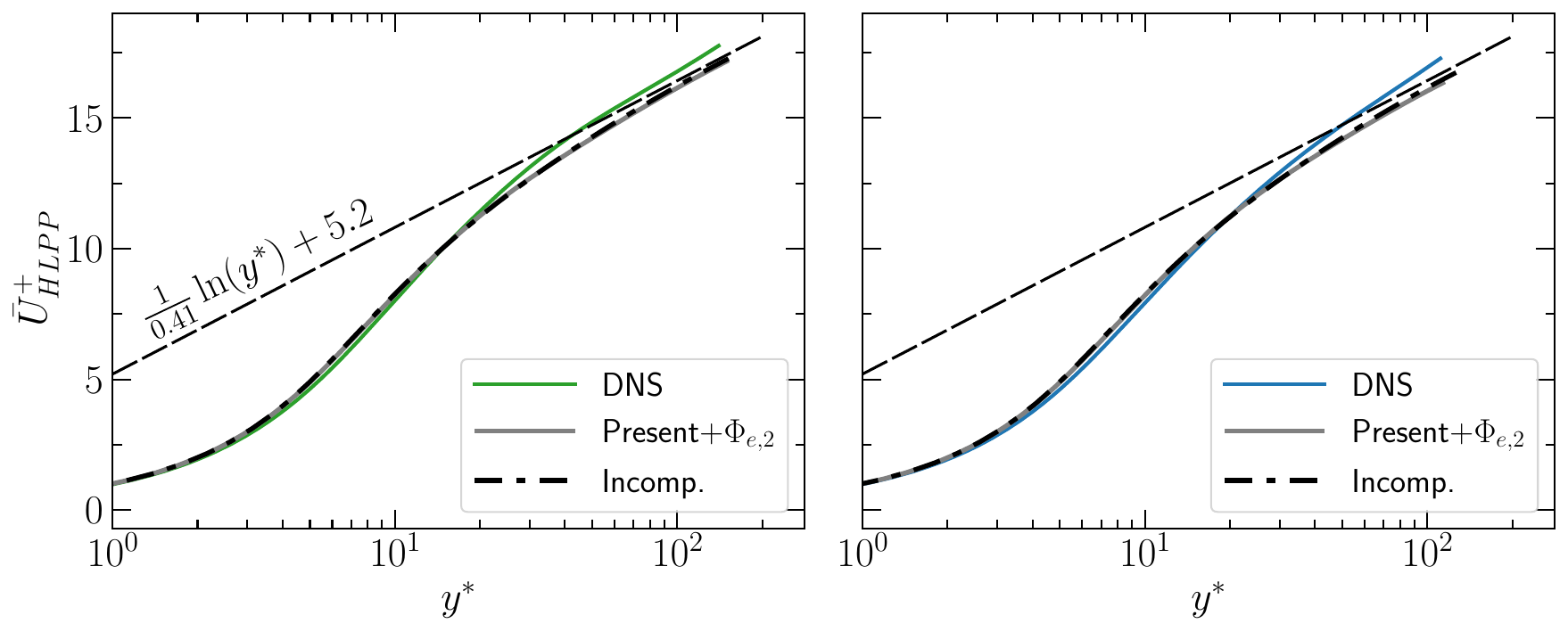}
\caption{{The HLPP-transformed~\eqref{eq:asif} mean velocity profiles as a function of the semi-local coordinate $y^*$, for the (left) $M_\infty=10.9$, $T_w/T_r=0.2$ \citep{huang2022direct} and (right) $M_\infty=13.64$, $T_w/T_r=0.18$ \citep{zhang2018direct} cases described in table~\ref{tab:casescfd}. These cases correspond to the fourth and fifth cases from left in figure~\ref{Fig:tblcfd}. The colored lines correspond to DNS, whereas the grey solid lines correspond to that estimated from the SST model with the present corrections, along with $\Phi_{e,2}$ in the energy equation. The dash-dotted black lines correspond to that estimated from the SST model for an incompressible (constant-property) case at similar $Re_\tau$ as the respective compressible cases described above.   
Note that the DNS profiles are plotted only until $y/\delta=0.2$ (edge of the inner layer) for clarity.}}
 \label{Fig:tblhlpp}
\end{figure}

While the temperature profiles improve, the velocity profiles slightly deteriorate when $\Phi_{e,2}$ is used as the source term. This is explained as follows. The present corrections (equations~\ref{Phi},~\ref{phi_om}~and~\ref{Dic_kom}) make the SST model results consistent with the HLPP transformation~\eqref{eq:asif}. In other words, the velocity profiles estimated by the corrected model, upon HLPP transformation, would collapse on to the velocity profile estimated by the model for an incompressible flow. 
Figure~\ref{Fig:tblhlpp} shows the HLPP-transformed profiles for the fourth and fifth cases (from left) in figure~\ref{Fig:tblcfd}. Clearly, the solid grey lines that correspond to the corrected model collapse on to the black dash-dotted lines that represent the SST model prediction for incompressible flows. However, these lines are shifted downwards with respect to the DNS (depicted by solid colored lines). Such a downward-shifted model profile, when inverse transformed with accurate mean properties would result in $\bar u^+$ which is also under-predicted with respect to the DNS. This is what we observed in figure~\ref{Fig:tblcfd} for these cases (compare solid grey and colored lines). On the contrary, with $\Phi_{e,1}$, due to inaccurate temperature, and hence, property profiles, there are error cancellations which results in more accurate $\bar u^+$ observed in figure~\ref{Fig:tblcfd} (compare long-dashed grey and solid colored lines).

\begin{figure}
	\centering	\includegraphics[width=\textwidth]{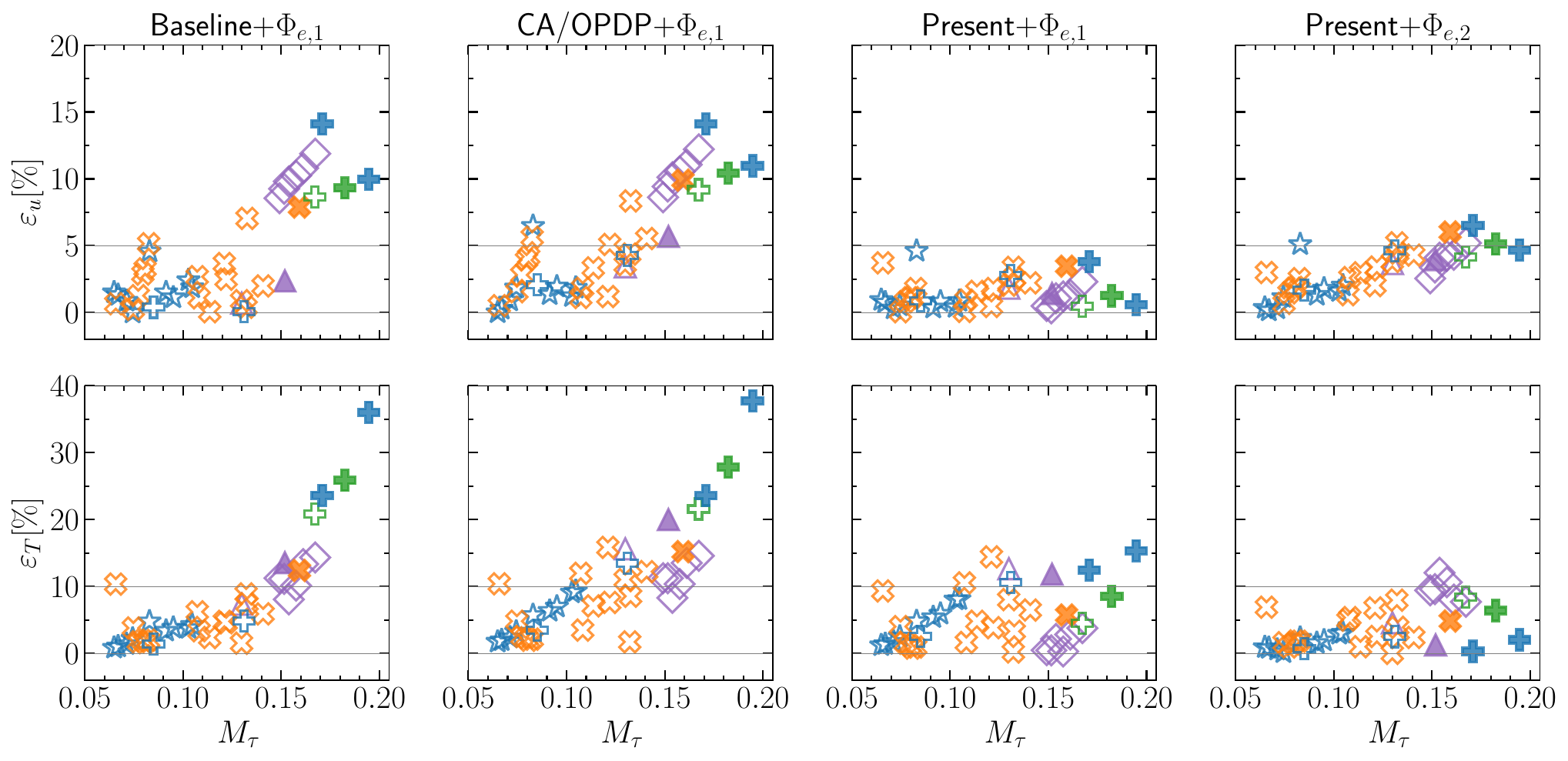}
\caption{{Percent error in velocity (top row) and temperature (bottom row) predictions for 39 compressible turbulent boundary layers from the literature as shown in table~\ref{tab:casescfd}. 
 The error is computed using equation~\eqref{errphi}.
  Symbols are as in table~\ref{tab:casescfd}. The filled symbols correspond to the cases whose velocity and temperature profiles are plotted in figure~\ref{Fig:tblcfd}. The gray horizontal lines in the inset indicate errors of 0\% and 5\% for velocity, and 0\% and 10\% for temperature.}}
 \label{Fig:tblerr}
\end{figure}

Next, we quantify the error in the velocity and temperature profiles in terms of the inner-layer edge values, namely 
\begin{equation}\label{errphi}
    \varepsilon_\phi =  \frac{\left| \phi_{y=0.2\delta} -  \phi_{y=0.2\delta}^{DNS}\right|}{\phi_{y=0.2\delta}^{DNS}} \times 100,
\end{equation}
where $\phi$ could either be $\bar u ^+$ or $\bar T/T_w$. Such an error definition represents the integrated or cumulative error in solving equation~\eqref{tbl:eqns} across the entire domain. In other words, it quantifies the overall accumulation of errors in key quantities such as \( \mu_t \), which is particularly relevant to the present study as our corrections directly impact \( \mu_t \).

Figure~\ref{Fig:tblerr} presents the error in velocity and temperature for the 39 boundary layer cases listed in table~\ref{tab:casescfd}. The titles of the subfigures correspond to various modeling corrections discussed earlier. With the baseline model (first column), consistent with the observations made in figure~\ref{Fig:tblcfd}, the errors in both velocity and temperature are significantly high reaching approximately 15\% and 40\%, respectively. When the CA/OPDP corrections are applied (second column), these errors remain similar or even slightly increase in some cases.  

With the implementation of the present inner-layer corrections (third column), the errors in both velocity and temperature reduce significantly, falling within approximately 5\% and 15\%, respectively. Finally, when the source term $\Phi_{e,1}$ is replaced with $\Phi_{e,2}$ (fourth column), the velocity errors slightly increase (as explained above), whereas the temperature errors improve across all cases except those of \cite{ceci2022numerical}. This discrepancy may be due to minor inaccuracies in the wall-cooling parameter ($B_q$) for these cases, which serves as an input to the solver \citep{jupnotebook}.

\subsubsection{Sensitivity of the results to $Pr_t$}
\begin{figure}
	\centering	\includegraphics[width=\textwidth]{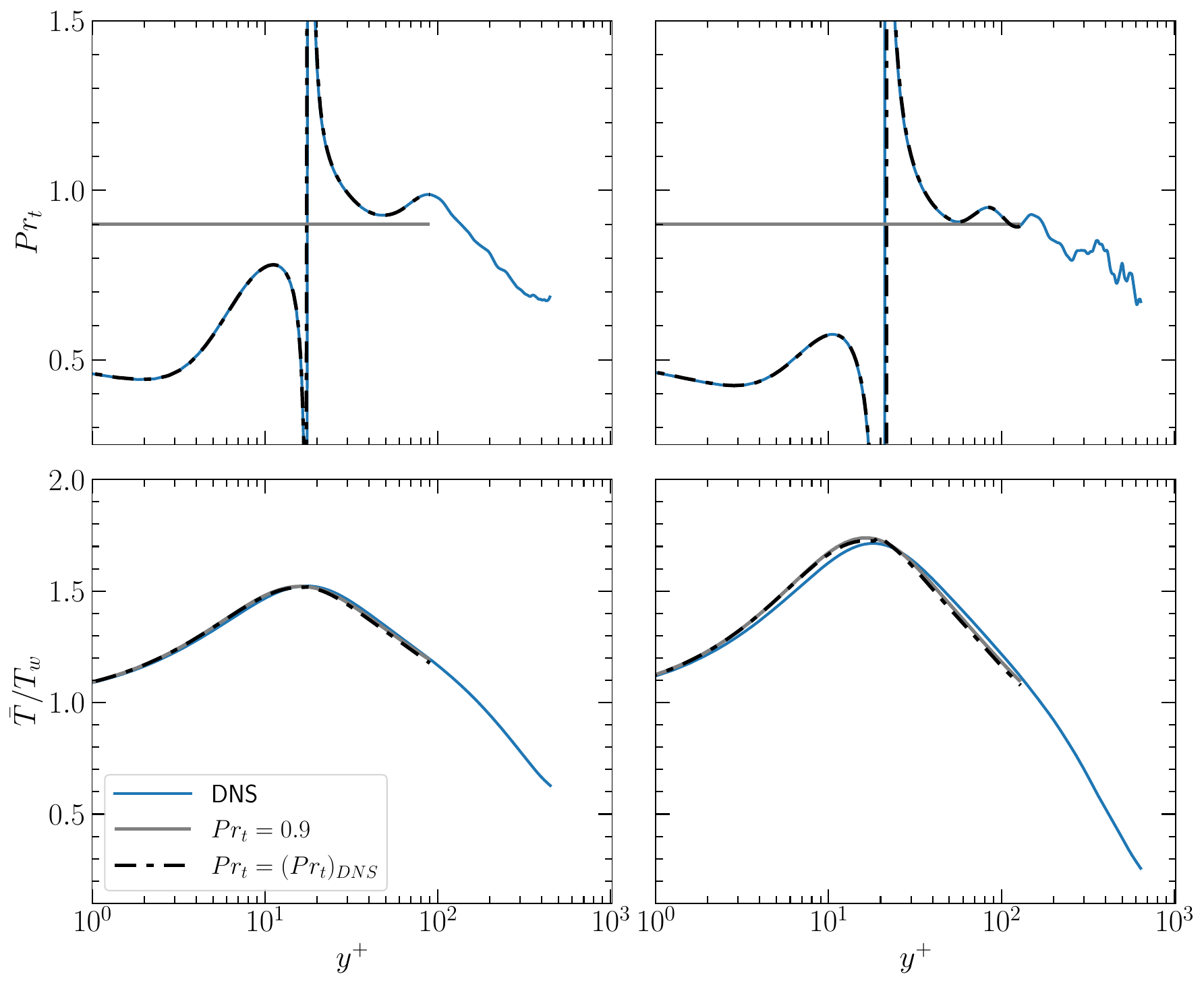}
\caption{{The turbulent Prandtl number ($Pr_t$; top row) and temperature profiles (bottom row) for the $M_\infty=5.84$, $T_w/T_r=0.25$ (left) and $M_\infty=13.64$, $T_w/T_r=0.18$ (right) cases of \citet{zhang2018direct}. These cases correspond to the third and fifth profiles in figure~\ref{Fig:tblcfd}. The solid colored lines represent the $Pr_t$ and $\bar{T}/T_w$ profiles extracted from the DNS. The solid grey lines correspond to the constant approximation $Pr_t=0.9$ in the top row and the resulting temperature profiles computed using this approximation in the bottom row. Finally, the dash-dotted black lines depict the DNS-based $Pr_t$ interpolated onto the solver’s mesh in the top row, and the corresponding temperature results obtained using this interpolated profile in the bottom row.}}
 \label{Fig:tblprt}
\end{figure}

The results discussed so far were obtained using a constant turbulent Prandtl number, $Pr_t = 0.9$, throughout the domain. In this section, we assess the sensitivity of the results to $Pr_t$ by comparing temperature profiles computed with $Pr_t$ from DNS to those obtained with $Pr_t = 0.9$. We continue using the present corrections, with $\Phi_{e,2}$ in the energy equation, and focus on the Mach 6 and 14 cases from \cite{zhang2018direct} (third and fifth cases from the left in figure~\ref{Fig:tblcfd}), as DNS data for $Pr_t$ is available for them.  

Figure~\ref{Fig:tblprt} (top row) compares the DNS-based $Pr_t$ (solid colored lines) with $Pr_t = 0.9$ (solid grey lines) for the Mach 6 (left) and Mach 14 (right) cases mentioned above. The difference is significant in the near-wall region ($y^* < 15$), but the profiles converge further away from the wall. To incorporate this DNS-based $Pr_t$ into our computations, we interpolate it onto the mesh used by our solver. The interpolated values closely match the DNS profiles, as seen from the dash-dotted black lines in figure~\ref{Fig:tblprt} (top row).  

Using these interpolated $Pr_t$ values, we recompute the temperature profiles and present them in figure~\ref{Fig:tblprt} (bottom row) with dash-dotted black lines. These results are nearly identical to those obtained with $Pr_t = 0.9$ (solid grey lines), suggesting that $Pr_t = 0.9$ is a reasonable approximation for the ideal-gas, air-like cases considered here.

The nearly identical results can be explained as follows. In the near-wall region, the molecular diffusion of energy dominates the turbulent diffusion. As a result, despite the significant mismatch between the DNS-based $Pr_t$ and $Pr_t=0.9$ in this region, the temperature profiles remain largely unaffected. Further away from the wall, where turbulent diffusion becomes significant, the DNS-based $Pr_t$ closely aligns with $Pr_t=0.9$, leading to similar temperature predictions. These findings contradict previous studies \citep{chen2024improved}, which state that the near-wall variations in the turbulent Prandtl number are the primary cause of errors in temperature predictions for high-speed flows.

\subsection{Fully developed channel flows}

\begin{figure}
	\centering	\includegraphics[width=\textwidth]{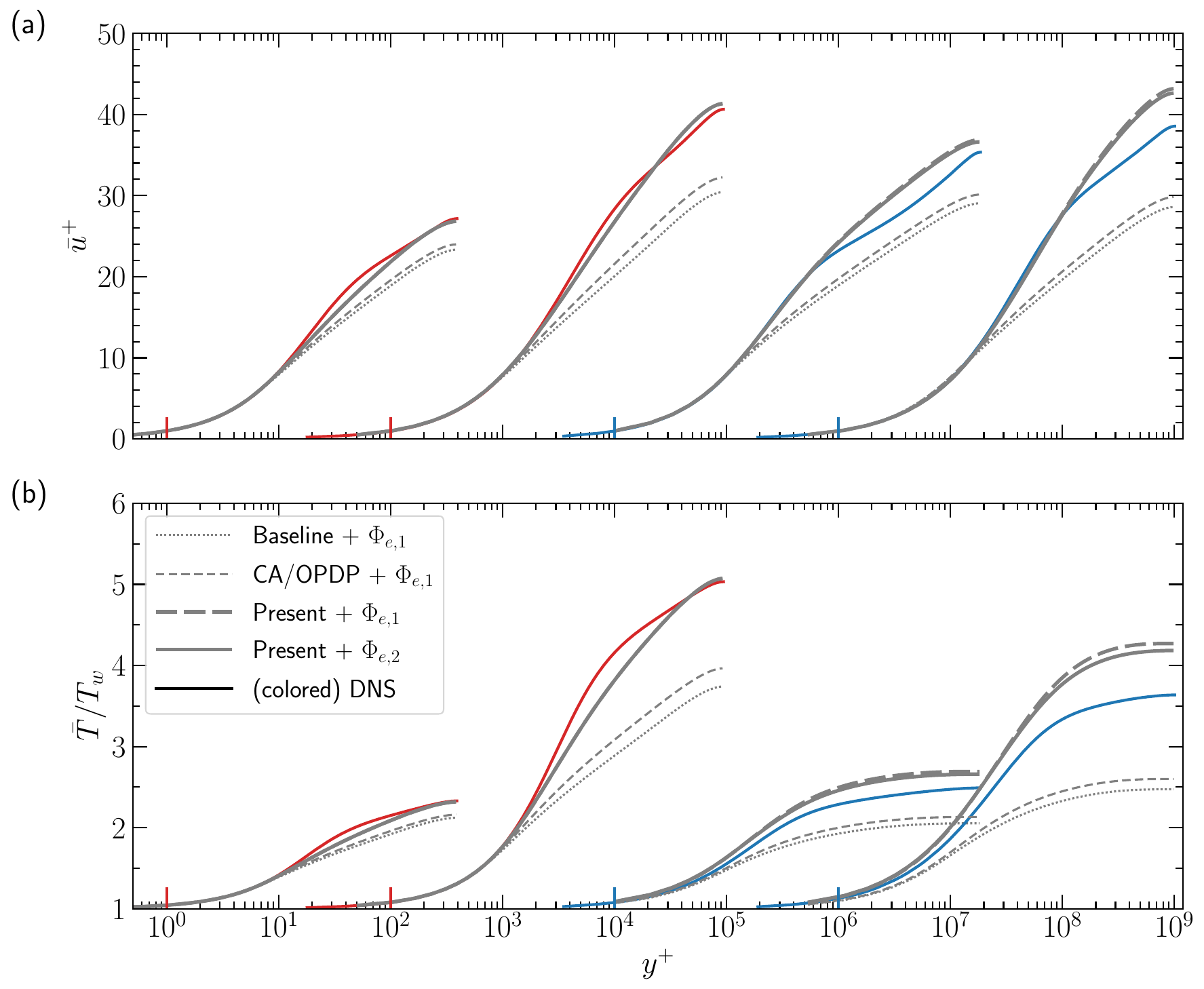}
	\caption{Computed mean velocity (a) and temperature (b) profiles compared to the DNS (colored solid lines) for the following turbulent channel flows: (left to right) $M_b = 0$, $Re_\tau=395$ (gas-like case of \cite{patel2015semi}); $M_b = 0$, $Re_\tau=950$ (gas-like case of \cite{pecnik2017scaling}); $M_b=3$, $Re_\tau = 1876$ and $M_b=4$, $Re_\tau = 1017$ (compressible cases of \cite{trettel2016mean}).
    The line colors match the color of the symbols for the respective authors reported in table~\ref{tab:casescfd}.
The line types in the legend are explained in the caption of figure~\ref{Fig:tblcfd}. Note that $\Phi_{e,1}$ and $\Phi_{e,2}$ only apply to the high-Mach number cases. For the low-Mach cases, the source term is a user-defined constant, as noted in section~\ref{implementation}.
For clarity, the velocity and temperature profiles for different cases are shifted by two decades along the abscissa. The colored vertical lines on the abscissa signify $y^+=10^0$ for each case, with their colors matching the corresponding cases.}
 \label{Fig:chancfd}
\end{figure}     

Figure~\ref{Fig:chancfd} shows the (a) mean velocity and (b) temperature profiles for two low-Mach (red) and two high-Mach-number turbulent channel flows (blue) described in the figure caption. The line types in the legend are similar to those discussed earlier for boundary layers, except that here $\Phi_{e,1}$ and $\Phi_{e,2}$ are only applicable to the high-Mach-number cases. For the low-Mach-number cases, the source term is a user-defined constant (see section~\ref{implementation}).

\textbf{Low-Mach-number cases} -- Let us first focus on the two low-Mach-number cases (first and second cases from the left).
While using the baseline model (grey dotted lines), both the velocity and temperature profiles are under-predicted compared to the DNS. 

Similar under-predictions are also observed when the CA/OPDP corrections are applied (see grey short-dashed lines). This inaccuracy originates from the fact that CA/OPDP corrections are based on Van Driest's scaling, which becomes inaccurate for strongly-cooled flows.

The grey long-dashed and the grey solid lines correspond to the results where the CA/OPDP corrections are replaced by the proposed variable-property corrections which account for variations in the viscous length scale (see equations~\ref{Phi}~and~\ref{phi_om}). These lines are equivalent for the low-Mach-number cases, and thus, only the grey solid lines are visible. This is because the main difference between these lines is in the source term ($\Phi_{e,1}$ and $\Phi_{e,2}$). For the low-Mach cases, these source terms are zero, since the Mach number is zero. 

Comparing the grey solid lines with the grey dotted and grey short-dashed lines, we note a substantial improvement for both velocity and temperature. This highlights the importance of accounting for viscous length scale variations in the model equations.

\textbf{High-Mach-number cases} --
Next, we look at the last two cases from the left, corresponding to the compressible turbulent channel flows. 
Similar to the low-Mach-number cases, the velocity and temperature profiles are under-predicted while using the baseline model with no corrections (see grey dotted lines) as well as while using the CA/OPDP corrections (see grey short-dashed lines). The results improve when the proposed corrections are used, as indicated by grey long-dashed lines, however, the velocity and temperature profiles are now over-predicted compared to the DNS. Changing the source term in the energy equation from $\Phi_{e,1}$ to $\Phi_{e,2}$ (grey solid lines) leads to little improvement, with the over-prediction largely unchanged.

\begin{figure}
	\centering	\includegraphics[width=\textwidth]{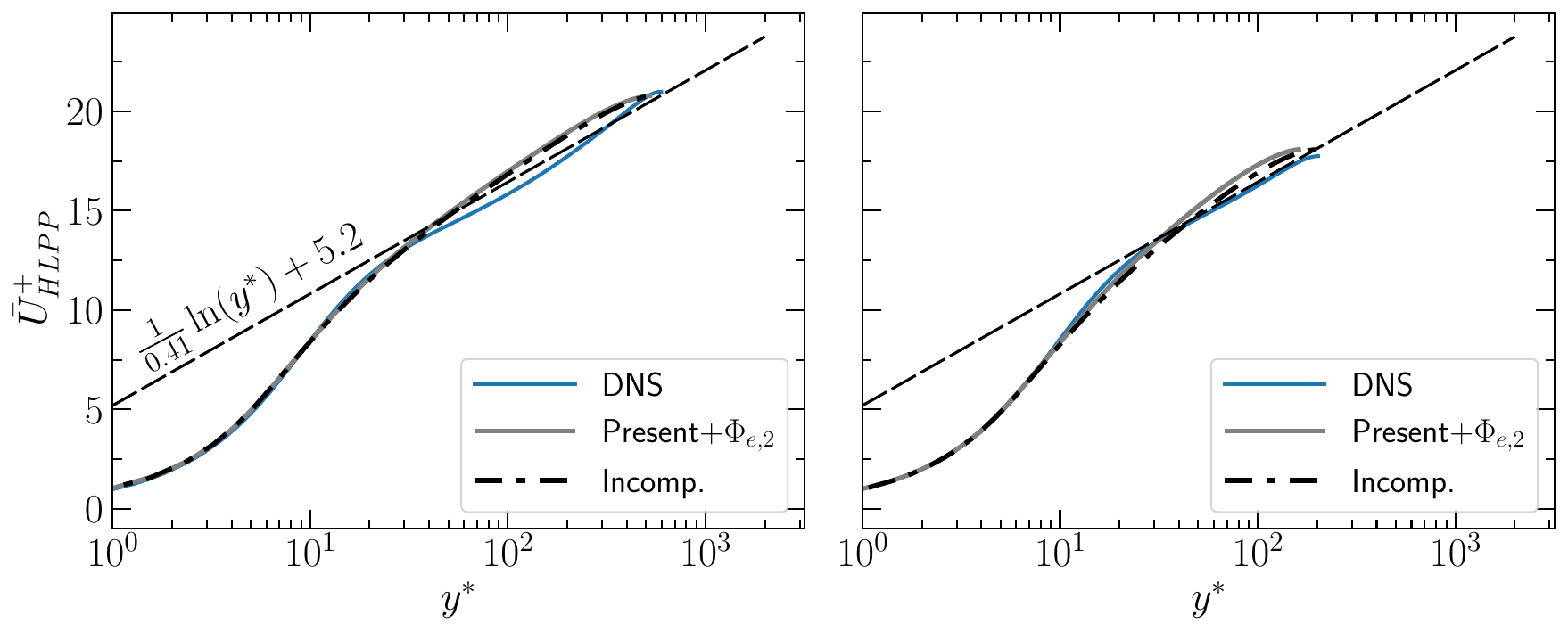}
\caption{{The HLPP-transformed~\eqref{eq:asif} mean velocity profiles as a function of the semi-local coordinate $y^*$, for the (left) $M_b=3$, $Re_\tau = 1876$ and (right) $M_b=4$, $Re_\tau = 1017$  cases of \cite{trettel2016mean} described in table~\ref{tab:casescfd}. These cases correspond to the third and fourth cases from left in figure~\ref{Fig:chancfd}. The colored lines correspond to DNS, whereas the grey solid lines correspond to that estimated from the SST model with the present corrections, along with $\Phi_{e,2}$ in the energy equation. The dash-dotted black lines correspond to that estimated from the SST model for an incompressible (constant-property) case at similar $Re_{\tau_c}^*$ (semi-local Reynolds number computed using channel centreline properties as $Re_{\tau_c}^* = \bar \rho_c u_{\tau_c}^* h/{\bar\mu_c}$) as the respective compressible cases described above.}}
 \label{Fig:chanhlpp}
\end{figure}

The possible reason for this over-prediction is discussed next. Previous studies have shown that HLPP-transformed DNS mean velocity profiles for compressible channel flows are shifted downwards in the logarithmic region compared to the incompressible law of the wall \citep{hasan2023incorporating}. This trend is also observed in figure~\ref{Fig:chanhlpp} for the Mach 3 case (left). However, for the Mach 4 case (right), such a downward shift is not evident, likely due to its low Reynolds number. (Specifically, the local Reynolds number in this case decreases from 1017 at the wall to below 250 at $y^*\approx15$, continuing to drop further away from the wall).
 
On the other hand, the transformed profiles predicted by the SST model slightly overshoot relative to the law of the wall, especially in the logarithmic region and beyond ($y^*>30$). 
As a result, these model profiles are consistently higher than the DNS in these regions, such that when inverse-transformed, the estimated $\bar u^+$ is also higher. The higher $\bar u^+$ directly influence the energy equation via the source terms $\Phi_{e,1}$ and $\Phi_{e,2}$, thereby leading to an over-prediction in temperature profiles. This, in turn, influences $\bar u^+$ through mean properties, further intensifying the over-prediction. 

Interestingly, when we break this coupling between velocity and temperature by using temperature from the DNS in our computations (not shown), we observe that the over-prediction in $\bar u^+$ is significantly reduced suggesting that the coupling between energy and momentum equations plays an important role in the higher $\bar u^+$ and $\bar T/T_w$ values.

\begin{figure}
	\centering	\includegraphics[width=\textwidth]{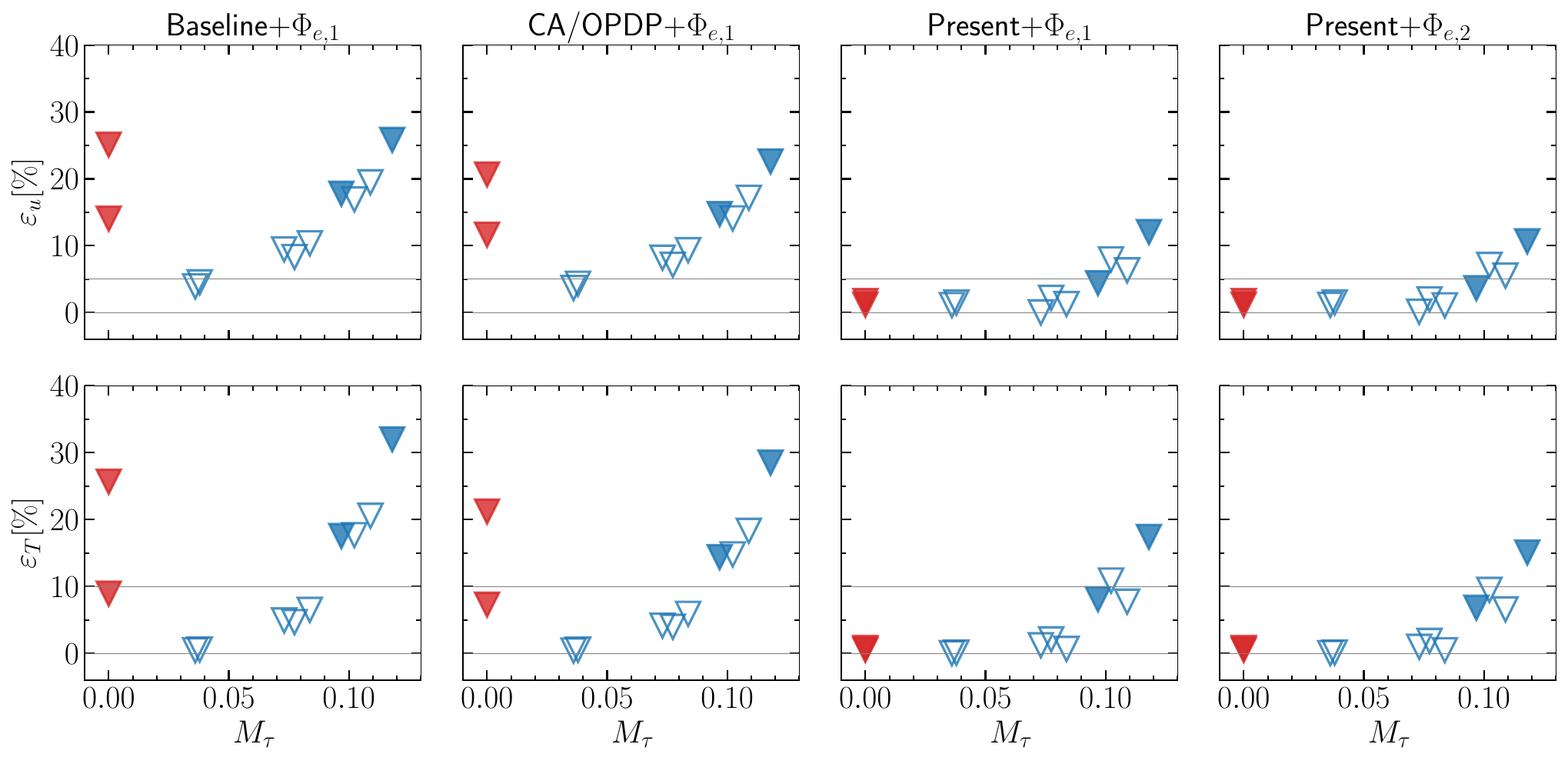}
\caption{{Percent error in velocity (top row) and temperature (bottom row) predictions for 11 turbulent channel flows from the literature as shown in table~\ref{tab:casescfd}. 
 The error is computed using equation~\eqref{errphi}.
  Symbols are as in table~\ref{tab:casescfd}. The filled symbols correspond to the cases whose velocity and temperature profiles are plotted in figure~\ref{Fig:chancfd}. The gray horizontal lines in the inset indicate errors of 0\% and 5\% for velocity, and 0\% and 10\% for temperature.}}
 \label{Fig:chanerr}
\end{figure}

Finally, we quantify the errors ($\varepsilon_\phi$) in the computed velocity and temperature profiles using equation~\eqref{errphi}, but now $\phi$ is taken at the channel centreline ($\phi_{y=h}$).

Figure~\ref{Fig:chancfd} shows the errors in the velocity profiles ($\varepsilon_u$) for two low-Mach-number and nine high-Mach-number channel flow cases listed in table~\ref{tab:casescfd}. Like in figure~\ref{Fig:tblerr}, the titles of the subfigures correspond to various modeling corrections discussed earlier. The errors using the baseline model with no corrections (first column) and with the CA/OPDP corrections (second column) remain similar, with the errors for velocity and temperature reaching up to approximately 25\% and 30\%, respectively. 

With our corrections (third and fourth columns), the errors in both velocity and temperature are significantly reduced, falling within approximately 10\% and 15\%, respectively. Furthermore, the errors remain similar for both $\Phi_{e,1}$ and $\Phi_{e,2}$, suggesting lower sensitivity of the channel flow results to the choice of the source term.

\section{\changed{Testing the proposed corrections on a two-dimensional boundary layer}}\label{full2d}

In the previous sections, we discussed the implementation and results for the inner layer of ZPG boundary layers and channel flows, that can be modeled using one-dimensional equations. However, it is so far unclear how the proposed corrections perform in a two-dimensional boundary layer simulation, where the outer layer is also solved for. 

To address this point, we performed a flat plate boundary layer simulation for the \(M_\infty = 10.9\), \(T_w/T_r = 0.2\) case of \cite{huang2022direct}, using {Ansys Fluent}. 
The computational domain spans \SI{1.5}{m} in the streamwise direction and \SI{0.2}{m} in the wall-normal direction, discretized using a structured mesh consisting of 1600 cells along the streamwise direction and 300 cells in the wall-normal direction. The mesh is stretched in the wall-normal direction such that the first grid point lies within one viscous wall unit from the wall. {With this mesh, we achieve a resolution of about $\Delta x^+ \approx 97 $ and $\Delta y^+_{\text{min}} \approx 0.014$ at the streamwise location whose profiles are shown in figure~\ref{Fig:tbl2d}. This resolution is finer than that adopted in \cite{danis2022compressibility} for the same case.}

A velocity-inlet boundary condition is applied at the inlet, prescribing a constant streamwise velocity of \SI{1778.4}{m/s}, pressure of \SI{1966.1}{Pa}, and temperature of \SI{64.4}{K}. At the wall, a no-slip boundary condition is imposed along with a constant wall temperature of \SI{300}{K}. A pressure far-field boundary condition is specified at the top boundary, with the pressure and temperature set to \SI{1966.1}{Pa} and \SI{64.4}{K}, matching their respective values at the inlet. At the outlet, a pressure-outlet boundary condition is used, prescribing the pressure to be equal to the free-stream pressure, and the backflow temperature is set to \SI{300}{K}. Note that the free-stream temperature specified in the DNS is \SI{66.5}{K}; here, we slightly reduce this value to account for the temperature jump across the oblique shockwave originating from the leading-edge, such that the post-shock temperature is approximately \SI{66.5}{K}. (The pressure also changes across the shockwave; however, the ratio of its post-shock value to the wall value remains unchanged. In contrast, because the wall temperature is fixed, the corresponding ratio for temperature is affected. This, in turn, alters the density ratio and ultimately influences the computed skin friction and heat transfer coefficients in figure~\ref{Fig:cfch}.)
With these boundary conditions, we perform the simulations using Fluent's density based solver. Further details on the simulation setup (for instance, the case file) can be found on our GitHub repository \citep{jupnotebook}. 
 
To implement the proposed corrections in Fluent, we use user-defined functions (UDFs). In these UDFs, we compute the source terms $\Phi_k$, $\Phi_\omega$ and $\Phi_{CD}$ and add them on the right-hand-side of the conventional SST model as indicated in equations~\eqref{fullsstk} and~\eqref{fullsstom}. We also redefine the eddy-viscosity as in equation~\eqref{sstic}. 
The reader is referred to our GitHub repository \citep{jupnotebook} for more details on the UDFs used to generate the results shown below.

To obtain the mean temperature profile, we solve the full energy equation:
\begin{align}\label{enfluent2a}
\frac{\partial \left[\bar\rho (e + \widetilde{u}_i\widetilde{u}_i + k)\right]}{\partial t}+\frac{\partial \left[\bar\rho \widetilde{u}_j (h + \widetilde{u}_i\widetilde{u}_i + k)\right]}{\partial x_j} =& \frac{\partial}{\partial x_j}\left[\left(\frac{\bar{\mu}c_p}{P r}+\frac{\mu_t c_p}{P r_t}\right) \frac{\partial \bar{T}}{\partial x_j}\right] + \nonumber\\
&\frac{\partial}{\partial x_j}\left[\widetilde{u}_i\left(2\left(\bar{\mu}+\mu_t\right) \widetilde{S}_{i j}\right)\right] + \nonumber\\
& \underbrace{\frac{\partial}{\partial x_j}\left[\left(\bar \mu+\sigma_{k} \mu_t\right) \frac{\partial k}{\partial x_j}\right] + \Phi_k}_{\text{variable-property-corrected (section~\ref{Sec:derivInner})}},
\end{align}
where $e$ and $h$ are the internal energy and enthalpy per unit mass, and $S_{ij}$ is the strain rate tensor. (Refer to section~\ref{sec:energyeq} for a discussion on the energy equation in the inner layer.) Adding $\Phi_k$ to the conventional equation implies that the viscous and turbulent diffusion of TKE are modeled after accounting for variable property effects, consistent with the diffusion term in the TKE transport equation (see section~\ref{Sec:derivInner}). 

Note that in Fluent, TKE is not included in the definition of total energy, and its diffusion does not appear in the energy equation. Therefore, to correctly solve equation~\eqref{enfluent2a}, it is necessary to add the TKE transport term \(\bar{\rho}\, Dk/Dt\) to the left-hand side of the energy equation and the TKE diffusion term to the right-hand side. We achieve this in Fluent by introducing a source term on the right-hand side of the energy equation, equal to the TKE diffusion term minus \(\bar{\rho}\, Dk/Dt\). This term is effectively equal to \(-P_k + \bar{\rho}\,\epsilon\), where \(P_k\) denotes the production of TKE. For further details, refer to the UDF provided in \citet{jupnotebook}.

In the inner layer, equation~\eqref{enfluent2a} reduces to equation~\eqref{energy7}. 
Interestingly, with $k$ taken from the SST model, solving equation~\eqref{enfluent2a} in the inner layer is equivalent to solving equation~\eqref{tbl:eqns} with $\Phi_e = \bar \mu (d \bar u/dy)^2 + \bar\rho\epsilon_{\text{sst}}$. 
However, as discussed above equation~\eqref{eps_epsw}, solving such an equation in the inner layer would lead to inaccuracies in predicting the mean temperature, since $\epsilon_{\text{sst}}$ tends to zero at the wall, instead of a non-zero value. Thus, there, we proposed replacing $\epsilon_{\text{sst}}$ with $\epsilon_{\text{eff}}$~\eqref{eps_epsw}, or in other words, we proposed adding a source term, namely, $\bar\rho\epsilon_{\text{eff}} - \bar\rho\epsilon_{\text{sst}}$ to $\bar \mu (d \bar u/dy)^2 + \bar\rho\epsilon_{\text{sst}}$, such that effectively we arrive at $\Phi_{e,2}$~\eqref{phie2}.

In the same spirit, to include the effects of $\epsilon_w$ in the two-dimensional solver, 
one needs to add a source term equal to $\bar\rho\epsilon_{\text{eff}} - \bar\rho\epsilon_{\text{sst}}$ on the right-hand-side of equation~\eqref{enfluent2a}. However, including such a source term does not lead to any improvement in the temperature profile (not shown), unlike what one would expect based on the one-dimensional results, where including the effects of $\epsilon_w$ improved the temperature profiles (see gray solid lines in figure~\ref{Fig:tblcfd}b). This is explained as follows: In the one-dimensional solver~\eqref{tbl:eqns}, the wall heat flux (first term on the right hand side) is fixed and provided as a boundary condition. Thus, any increase in $\Phi_e$ resulting from adding the source term $\bar\rho\epsilon_{\text{eff}} - \bar\rho\epsilon_{\text{sst}}$ would directly reduce the conductive heat flux, thereby reducing the temperature peak. Contrastingly, in the two-dimensional solver, the wall temperature is fixed, and the heat flux is allowed to vary. Now, since the integral of the added source term from the wall to the free-stream at a particular streamwise location is non-zero, it directly results in higher wall heat flux, rather than a lower conductive flux close to the wall. Thus, the temperature profile does not improve. Providing an accurate source term using the SST model, without erroneously adding to the wall heat flux, still remains unclear. Nevertheless, here, we provide the results using equation~\eqref{enfluent2a}, i.e., without accounting for the influence of $\epsilon_w$.

\begin{figure}
	\centering	\includegraphics[width=\textwidth]{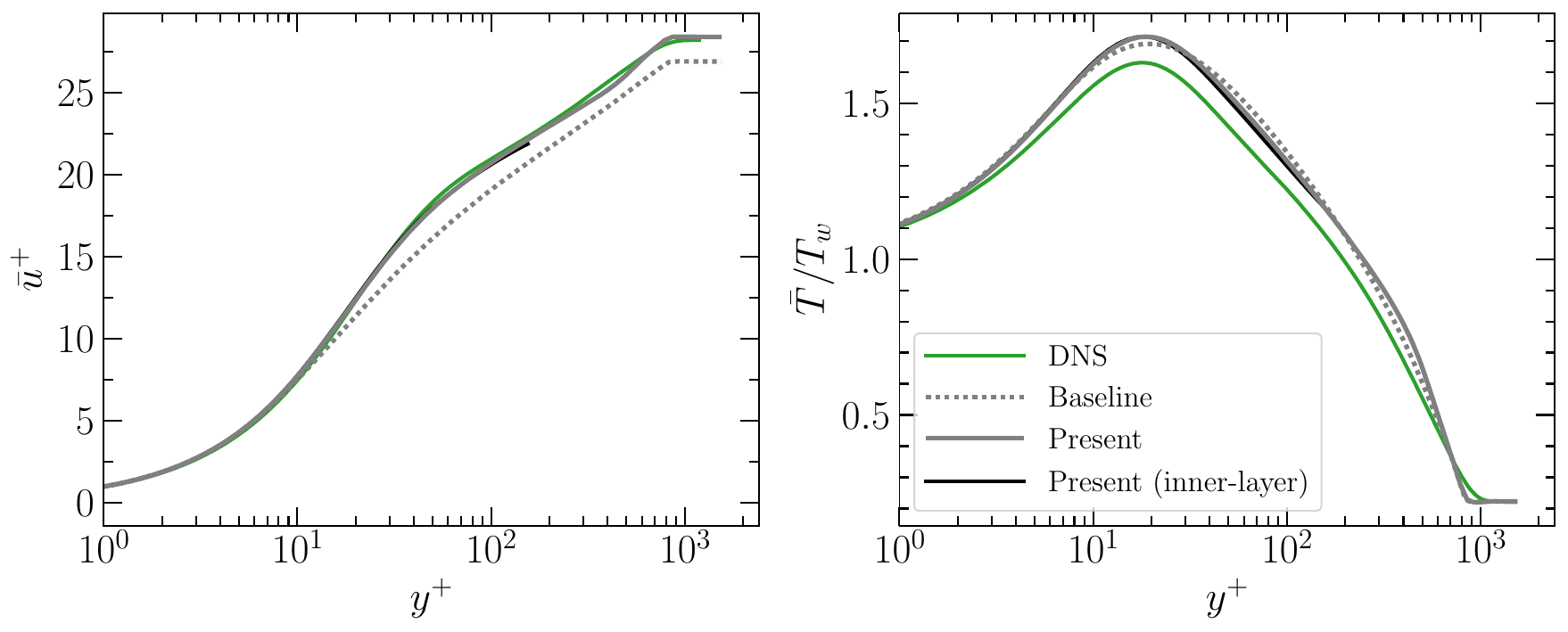}
\caption{Computed full mean velocity (left) and temperature (right) profiles compared to the DNS (colored solid lines) for the following turbulent boundary layer: $M_\infty=10.9$, $T_w/T_r=0.2$ $Re_\tau=774$ \citep{huang2022direct}. 
Refer to the legend for line types. `Baseline' represents the SST model without corrections, while `Present' signifies the corrections proposed in this paper (equations~\ref{blend1},~\ref{blend2},~\ref{phicd}~and~\ref{Dic_kom}). 
Note that, in contrast with figure~\ref{Fig:tblcfd}, where only the inner layer was solved, here both the baseline and present results are obtained by solving the entire boundary layer.
Also note that these results are computed with~\eqref{enfluent2a} as the energy equation, except that for the baseline case $\Phi_k=0$.
The solid black lines correspond to the profiles obtained after solving the inner-layer equations~\eqref{tbl:eqns} with $\Phi_e = \bar \mu (d\bar u/dy)^2 + \bar\rho \epsilon_{\text{sst}}$. 
}
 \label{Fig:tbl2d}
\end{figure}

Figure~\ref{Fig:tbl2d} shows the mean velocity (left) and temperature (right) profiles for the $M_\infty=10.9$ case described above. These profiles are extracted from the streamwise location where the friction Reynolds number matches the DNS value, i.e., $Re_\tau\approx774$. 
The solid colored lines represent the DNS profiles, while the dotted gray lines depict the predictions from the conventional SST model. The solid gray lines show the results obtained using the corrected SST model equations~\eqref{fullsstk} and~\eqref{fullsstom}, along with the modified eddy viscosity~\eqref{sstic}.
The solid black lines correspond to the results obtained after solving the one-dimensional equations with the present inner-layer corrections and with $\Phi_e = \bar \mu (d \bar u/dy)^2 + \bar\rho\epsilon_{\text{sst}}$ (see section~\ref{impl_bl}). This $\Phi_e$ is consistent with equation~\eqref{enfluent2a} in the inner layer, as discussed earlier.    

As seen in figure~\ref{Fig:tbl2d}, including the proposed corrections significantly improves the velocity profile compared to the one obtained with the baseline SST model (compare the dotted and solid gray lines). This improvement mainly originates from the inner-layer corrections. We confirmed this by setting $\Phi_k^{\textrm{out}}$, $\Phi_\omega^{\textrm{out}}$ and $\Phi_{CD}$ equal to zero in our simulations, such that we solve the baseline model in the outer layer. The resulting profiles were similar to those obtained when the outer-layer corrections were included. 
In contrast to the velocity, the temperature profiles remain largely unchanged.

Figure~\ref{Fig:tbl2d} also shows that, in the inner layer, the velocity and temperature profiles computed using Fluent (gray solid line) collapses well on to the respective profiles computed using the one-dimensional approach discussed earlier (black solid lines). 
Note that in Fluent, the molecular Prandtl number for air is close to 0.74, and the turbulent Prandtl number is assumed to be 0.85. Thus, to be consistent, the one-dimensional results are also computed with these $Pr$ and $Pr_t$ values.   

\begin{figure}
	\centering	\includegraphics[width=\textwidth]{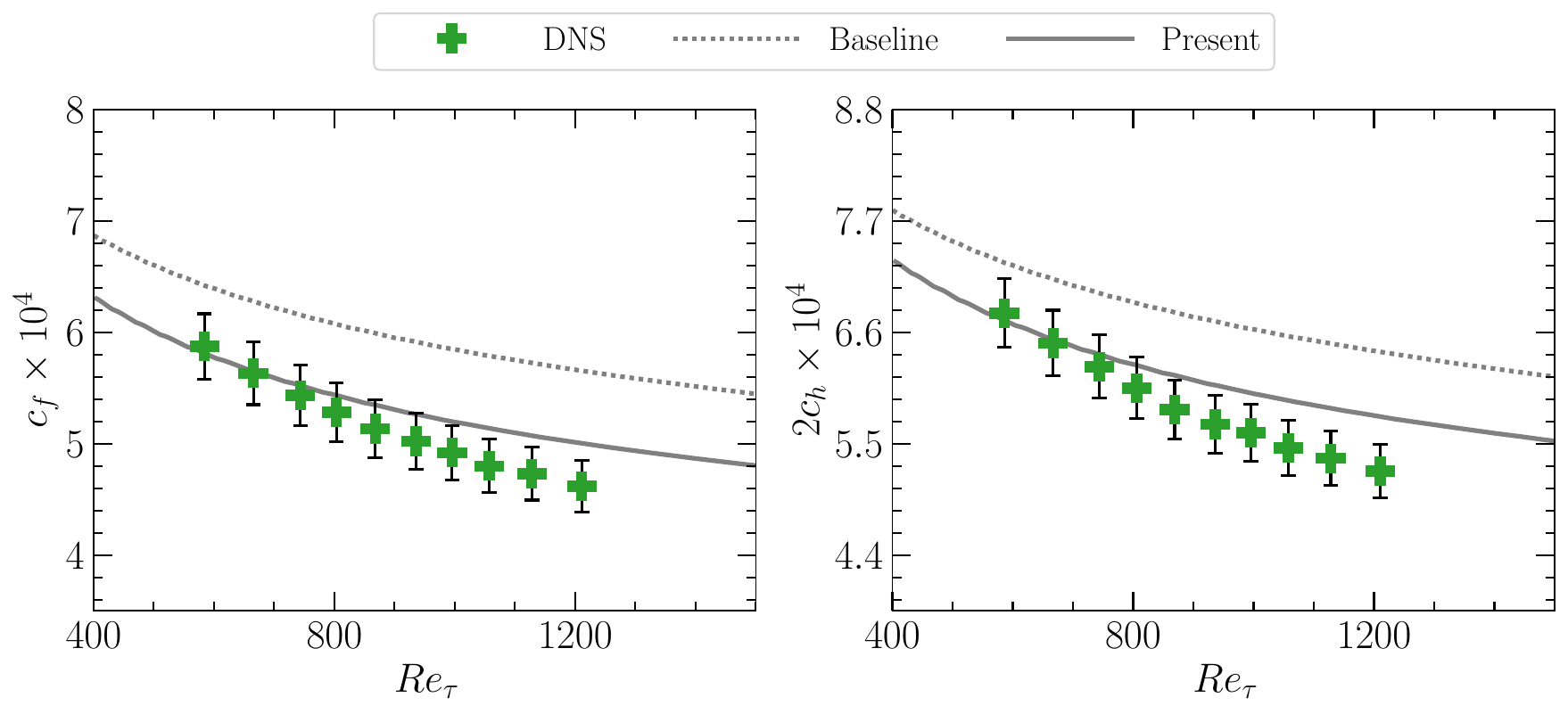}
\caption{The skin-friction (left) and heat transfer coefficients (right) compared to the DNS (symbols) for the $M_\infty=10.9$, $T_w/T_r=0.2$ turbulent boundary layer of \cite{huang2022direct}. 
The dotted and solid gray lines correspond to the quantities obtained with the uncorrected and corrected SST models, respectively.
The black vertical lines indicate an error margin of +/-5\%. Note that in the right subfigure, the \(y\)-axis has been scaled by a factor of 1.1, corresponding to the assumed value of the ratio \(2\,c_h / c_f\).
}
 \label{Fig:cfch}
\end{figure}

Finally, in figure~\ref{Fig:cfch}, we present the distribution of the skin-friction ($c_f = \tau_w/[0.5 \rho_\infty U_\infty^2]$) and heat transfer ($c_h = q_w/[\rho_\infty U_\infty c_p (T_r-T_w)]$) coefficients along the plate, as a function of $Re_\tau$. The filled symbols represent the DNS data of \cite{huang2022direct}, and the black vertical lines represent an error margin of +/-5\%. 
Clearly, the $c_f$ and $c_h$ predictions obtained from the corrected SST model are closer to the DNS values than those obtained from the conventional model (compare the dotted and solid gray lines). 

\section{Conclusion}
We have presented a novel approach to derive compressibility corrections for turbulence models. Using this approach, we have derived variable-property corrections that take into account the different scaling characteristics of the inner and outer layers.
In addition, we have formulated corrections that account for the change in near-wall damping of turbulence {due to intrinsic compressibility effects}. We have also tested different approximations for the source terms in the energy equation, as well as the assumption of a constant turbulent Prandtl number ($Pr_t$). Our findings, based on the $k$-$\omega$ SST model, are summarized below.

The proposed inner-layer corrections when compared {with the baseline model (with no corrections) and} with the state-of-the-art CA/OPDP corrections \citep{catris2000density, pecnik2017scaling, otero2018turbulence} produce significantly more accurate results for both turbulent boundary layers and channel flows.

In the context of temperature prediction, we highlighted the importance of {accurately modeling the source term} in the energy equation. When combined with the proposed compressibility corrections, these source terms substantially improve the accuracy of the peak temperature in cooled-wall turbulent boundary layers.

Another key factor in temperature prediction is the turbulent Prandtl number ($Pr_t$). By comparing two temperature profile predictions  -- one using a constant $Pr_t$ of 0.9 and the other using a $Pr_t$ distribution obtained from existing DNS data -- we find the results to be nearly identical (tested for a $M_\infty=6$ and $M_\infty=14$ boundary layer). This suggests that a constant $Pr_t$ of 0.9 is a reasonable approximation, {contradicting previous studies that identify the inaccuracies in the turbulent Prandtl number to be the primary source of error.}

\changed{Finally, we tested the proposed inner- and outer-layer corrections across the entire boundary layer using Ansys Fluent. Compared to the baseline model, the corrections significantly improve the accuracy of the mean velocity profile. However, the temperature profile remains largely unchanged. 
Moreover, the corrections also improve the accuracy of the predicted skin friction and heat transfer coefficients.
}

We recommend the following for future studies.
(1) \changed{The formulation of the corrections in cases where the wall-normal direction does not align with the coordinate axis (see Appendix~\ref{app:normal}) should be tested.}
(2) The performance of the proposed variable-property corrections 
should be tested in flows involving complex geometries, especially where the determination of the wall-normal direction is challenging (e.g. flows near corners).
(3) Also, the efficacy of the damping function~\eqref{Dic_kom} in more complex flow configurations must be tested.
(4) \changed{The possibility of adding an external heat source term to the full energy equation~\eqref{enfluent2a}, without erroneously adding to the wall heat flux, must be explored.}
(5) Using the same approach as presented in this paper for the SST model, the compressibility corrections for other turbulence models should be derived and tested. In Appendix~\ref{Sec:appSA}, we present and test the corrections for the Spalart-Allmaras (SA) model.
(6) For the SA model, improved alternatives for the TKE dissipation rate are needed to achieve even more accurate temperature profile predictions. \changed{It should also be tested whether such alternatives could improve temperature predictions using mixing-length models, for example in wall-modeled large eddy simulations (WMLES).}
(7) \changed{The influence of including $\Phi_k$ in the energy equation~\eqref{enfluent2a} should be explored using a turbulence model which predicts non-zero TKE diffusion and dissipation values at the wall (e.g. $v^2$-$f$ model \citep{durbin1991near}).}
\appendix

\section{{Deriving the outer-layer corrections}}\label{Sec:appouter}

\begin{table}
\centering\renewcommand{\arraystretch}{1.5}
\begin{tabular}{m{2.4cm} m{5cm} m{5cm} }
Quantity &  Incompressible (classical) & Compressible (semi-local) 
\\ \hline
Wall distance & $y^\circledplus = {y}/{\delta}$ & $y^\circledast = {y}/{\delta}$
\\
Mean shear & ${d\bar u^\circledplus}/{d y^\circledplus} = ({\delta}/{u_\tau}){d\bar u}/{d y} $ & ${d\bar u^{\circledast}}/{d y^\circledast} = ({\delta}/{u_\tau^*}){d\bar u}/{d y}$ 
\\ 
TKE & $k^\circledplus = {k}/{u_\tau^2}$ & $k^\circledast = {k}/{{u_\tau^*}^2}$ 
\\ 
{Turb. diss.} & {$\epsilon^\circledplus = {\epsilon}/({{u_\tau}^3/\delta})$} & {$\epsilon^\circledast = {\epsilon}/({{u_\tau^*}^3/\delta})$} 
\\
Spec. turb. diss. & $\omega^\circledplus = {\omega}/({u_\tau/\delta})$ & $\omega^\circledast = {\omega}/({u_\tau^*/\delta})$ 
\\
Eddy visc. & $\mu_t^\circledplus = {\mu_t}/({\rho_w u_\tau \delta}) = {(\mu_t/\mu_w)}/{Re_\tau}$ & $\mu_t^\circledast = {\mu_t}/({\bar\rho u_\tau^* \delta}) = {(\mu_t/\bar \mu)}/{Re_\tau^*}$ 
\\
Dyn. visc. & $\mu^\circledplus = {\mu_w}/({\rho_w u_\tau \delta}) = {1}/{Re_\tau}$ & $\bar \mu^\circledast = {\bar \mu}/({\bar\rho u_\tau^* \delta}) = {1}/{Re_\tau^*}$ 
\end{tabular}
\captionof{table}{An example of quantities that are classically and semi-locally scaled in the outer layer. The superscript `$\circledplus$' indicates classical outer-layer scaling, where as the superscript `$\circledast$' signifies semi-local outer-layer scaling. Note that $\bar u^\circledast$ in ${d\bar u^\circledast}/{d y^\circledast}$ represents the Van Driest transformed mean velocity, as defined in~\eqref{vdtrans}.}
\label{scaletabouter}
\renewcommand{\arraystretch}{1.0}
\end{table}

In this section, we derive the outer-layer corrections based on the logarithmic region. These corrections are then extended to the defect layer under the assumption that the advection terms remain unaffected by variable-property corrections. This approach was also employed in \cite{catris2000density}.

We start by writing the the modeled turbulence kinetic energy equation in the logarithmic region of a canonical incompressible {constant-property} flow as
\begin{equation}
    \mu_t \left(\frac{d\bar u}{dy}\right)^2 - \beta^\star\rho_w k \omega + \frac{d}{dy}\left[\left(\mu_w + \sigma_k \mu_t\right) \frac{d k}{dy}  \right] = 0.
\end{equation}
(Note that we do not drop the diffusion term in the log region, since it may become relevant in the defect layer where the derived corrections will be extended. Also, the viscous diffusion term, despite being irrelevant in the outer layer is retained for the sake of completeness).

Rewriting this equation using the non-dimensional form of the variables, as given in table~\ref{scaletabouter}, leads to
\begin{equation}\label{incomprefeqnout}
    {\mu_t^\circledplus} \left(\frac{d\bar u^\circledplus}{dy^\circledplus}\right)^2 - \beta^\star k^\circledplus \omega^\circledplus + \frac{d}{dy^\circledplus}\left[\left(1 + \sigma_k \mu_t^\circledplus\right) \frac{d k^\circledplus}{dy^\circledplus}  \right] = 0,
\end{equation}
where the superscript `$\circledplus$' denotes the classical scaling in the outer layer. Now, following section~\ref{Sec:derivInner}, we replace all classically scaled variables with their semi-locally scaled counterparts (refer table~\ref{scaletabouter}), which gives 
\begin{equation}\label{tkestarout}
    {\mu_t^\circledast} \left(\frac{d\bar u^\circledast}{dy^\circledast}\right)^2 - \beta^\star k^\circledast \omega^\circledast + \frac{d}{dy^\circledast}\left[\left(1 + \sigma_k \mu_t^\circledast\right) \frac{d k^\circledast}{dy^\circledast}  \right] = 0,
\end{equation}
where the superscript `$\circledast$' denotes semi-local scaling in the outer layer. Rewriting equation~\eqref{tkestarout} using the dimensional form of the variables (see table~\ref{scaletabouter}), we get
\begin{equation}\label{tkestar2out}
    \frac{\mu_t}{\bar \rho u_\tau^* \delta} \left(\frac{\delta}{u_\tau^*}\right)^2\left(\frac{d\bar u}{dy}\right)^2 - \beta^\star \frac{k}{{u_\tau^*}^2} \frac{\omega}{u_\tau^*/\delta} + \frac{d}{d(y/\delta)}\left[\left(\frac{\bar \mu}{\bar \rho u_\tau^* \delta} + \sigma_k \frac{\mu_t}{\bar \rho u_\tau^* \delta}\right) \frac{d (k/{u_\tau^*}^2)}{d(y/\delta)}  \right] = 0.
\end{equation}
Now, using the definition of $u_\tau^*$ from equation~\eqref{Eq.vd,mork}, we get
\begin{equation}\label{tkedim1out}
    \frac{\mu_t}{\sqrt{\tau_w \bar \rho}} \frac{\delta}{\tau_w/\bar\rho}\left(\frac{d\bar u}{dy}\right)^2 - \beta^\star \frac{\bar \rho}{\tau_w} k \frac{\delta}{\sqrt{\tau_w/\bar\rho}}\omega + \frac{d}{d(y/\delta)}\left[\left(\frac{\bar \mu}{\sqrt{\tau_w \bar\rho}\delta} + \sigma_k \frac{\mu_t}{\sqrt{\tau_w \bar\rho}\delta}\right) \frac{d (\bar \rho k/\tau_w)}{d(y/\delta)}  \right] = 0,
\end{equation}
which when divided by $\sqrt{\bar \rho} \delta/\tau_w^{1.5}$ gives
\begin{equation}\label{tkedim2out}
    {\mu_t}\left(\frac{d\bar u}{dy}\right)^2 - \beta^\star {\bar \rho}k \omega + \frac{1}{\sqrt{\bar \rho}}\frac{d}{dy}\left[\left(\bar \mu + \sigma_k {\mu_t}\right)\frac{1}{\sqrt{\bar \rho}} \frac{d (\bar \rho k)}{dy}  \right] = 0,
\end{equation}
These corrections are identical to the ones proposed by \cite{otero2018turbulence} for the entire boundary layer.

Now, as done in section~\ref{Sec:derivInner}, we express these corrections in the form of a source term as
\begin{equation}\label{phikoutout}
    \Phi_k^{\textrm{out}}  =  \frac{1}{\sqrt{\bar \rho}}\frac{\partial}{\partial y}\left[\left(\bar \mu+\sigma_{k} \mu_t \right)\frac{1}{\sqrt{\bar \rho}} \frac{\partial (\bar \rho k)}{\partial y}  \right] - \frac{\partial}{\partial y}\left[\left(\bar \mu+\sigma_{k} \mu_t\right) \frac{\partial k}{\partial y}\right].
\end{equation}
Note that partial derivatives are used because derivatives in the wall-parallel directions are non-zero in the outer layer.   

Repeating the same procedure for the $\omega$ equation, we get
\begin{equation}\label{phiomoutout}
\Phi_\omega^{\textrm{out}}  =  \frac{\partial}{\partial y}\left[\left(\bar \mu+\sigma_{\omega} \mu_t \right)\frac{1}{\sqrt{\bar \rho}} \frac{\partial (\sqrt{\bar \rho} \omega)}{\partial y}  \right] - \frac{\partial}{\partial y}\left[\left(\bar \mu+\sigma_{\omega} \mu_t\right) \frac{\partial \omega}{\partial y}\right].
\end{equation}
\changed{These source terms are blended with their inner-layer counterparts and then used in the full $k$ and $\omega$ transport equations as discussed in section~\ref{Sec:appblend}.}

\section{\changed{Formulation of the corrections in a general coordinate system}}\label{app:normal}

In section~\ref{Sec:derivvp}, we derived the corrections for the case in which the wall-normal direction was aligned with the $y$-axis. For arbitrary wall orientations, the source terms, particularly in the inner layer, must be modified. 
In such flows, we need to include the following source term in the inner layer:
\begin{eqnarray}\label{eqn1}
  \Phi_k^{\textrm{in}} = \frac{S_n}{\bar \mu}\frac{\partial}{\partial n }\left[\mu_\text{eff}\frac{S_n}{\bar \mu} \frac{\partial (\bar \rho k)}{\partial n}  \right] - \frac{\partial}{\partial n}\left[\mu_\text{eff}\frac{\partial k}{\partial n}  \right],  
\end{eqnarray}
where $n$ is the wall-normal coordinate and $\mu_\text{eff} = \bar \mu + \sigma_k {\mu_t}$, and where
\begin{equation}\label{Sn}
    S_n = \left(\frac{\partial \left(\ell\sqrt{\bar\rho}/\bar\mu\right)}{\partial n}\right)^{-1}=\left(\frac{\sqrt{\bar\rho}}{\bar\mu} + \ell \,\frac{\partial \left(\sqrt{\bar\rho}/\bar\mu\right)}{\partial n}\right)^{-1}.
\end{equation}
To further simplify this term, we note that the conventional diffusion term is coordinate independent. For a two-dimensional system, this implies that
\begin{equation}
  \frac{\partial}{\partial n}\left[\mu_\text{eff}\frac{\partial k}{\partial n}  \right] + \frac{\partial}{\partial t}\left[\mu_\text{eff}\frac{\partial k}{\partial t}  \right] =     \frac{\partial}{\partial x}\left[\mu_\text{eff}\frac{\partial k}{\partial x}  \right] 
  + \frac{\partial}{\partial y}\left[\mu_\text{eff}\frac{\partial k}{\partial y}  \right],  
\end{equation}
where $t$ is the wall-parallel direction, and where $x$ and $y$ are the generic cartesian coordinates. Close to the wall, the diffusion term is mainly dominated by the wall-normal component. Thus, one could neglect the second term on the left-hand side to get 
\begin{equation}\label{dkdy}
  \frac{\partial}{\partial n}\left[\mu_\text{eff}\frac{\partial k}{\partial n}  \right] \approx     \frac{\partial}{\partial x_j}\left[\mu_\text{eff}\frac{\partial k}{\partial x_j}  \right].  
\end{equation}
Similarly, we can write
\begin{equation}\label{dkdy2}
  \frac{\partial}{\partial n}\left[\mu_\text{eff}\frac{S_n}{\bar \mu} \frac{\partial (\bar\rho k)}{\partial n}  \right] \approx     \frac{\partial}{\partial x_j}\left[\mu_\text{eff}\frac{S_n}{\bar \mu}\frac{\partial (\bar\rho k)}{\partial x_j}  \right],  
\end{equation}
where $\mu_\text{eff} S_n/\bar\mu$ is the total diffusion coefficient, and $\bar \rho k$ is the diffused quantity.

Replacing the wall normal diffusion terms in equation~\eqref{eqn1} using equations~\eqref{dkdy} and~\eqref{dkdy2}, we get 
\begin{eqnarray}\label{eqnfinal}
  \Phi_k^{\textrm{in}} = \frac{S_n}{\bar \mu}\frac{\partial}{\partial x_j }\left[\mu_\text{eff}\frac{S_n}{\bar \mu} \frac{\partial (\bar \rho k)}{\partial x_j}  \right] - \frac{\partial}{\partial x_j}\left[\mu_\text{eff}\frac{\partial k}{\partial x_j}  \right],  
\end{eqnarray}
where $S_n$ can be expressed using the general coordinate system as
\begin{eqnarray}\label{snfinal}
        S_n = \left(\frac{\sqrt{\bar\rho}}{\bar\mu} + \ell \,n_j\frac{\partial \left(\sqrt{\bar\rho}/\bar\mu\right)}{\partial x_j} \right)^{-1},
\end{eqnarray}
where $n_j$ represents a unit vector in the wall-normal direction. One could follow a similar approach to derive $\Phi_\omega^{\text{in}}$.

In the outer layer, we propose applying the corrections equivalently in all the directions. For instance,
\begin{eqnarray}
\Phi_k^{\textrm{out}}  =  \frac{1}{\sqrt{\bar \rho}}\frac{\partial}{\partial x_j}\left[\mu_\text{eff}\frac{1}{\sqrt{\bar \rho}} \frac{\partial (\bar \rho k)}{\partial x_j}  \right] - \frac{\partial}{\partial x_j}\left[\mu_\text{eff} \frac{\partial k}{\partial x_j}\right],
\end{eqnarray}
and similarly for $\Phi_\omega^{\text{out}}$
and $\Phi_{CD}$.

\section{{The formulation and tuning of $f(M_t)$}}\label{Sec:appfmt}

The first step in obtaining the function $f(M_t)$ is to determine the value of $K$ in equation~\eqref{dampkom}. A simple approach to set $K$ is based on the value of $R_t$ for the turbulence model at $y^* \approx 17$. The value $y^* = 17$ corresponds to $A^+$ in the mixing-length damping function, as described in equation~\eqref{damp}.

Next, we treat $f(M_t)$ to be uniform in the domain (constant) and gradually increase its value from 0 in the denominator of equation~\eqref{dampkom}. Simultaneously, we solve the inner-layer equations discussed in section~\ref{impl_bl}, where the eddy viscosity formulation is multiplied with the damping function given in equation~\eqref{dampkom}. For these calculations, we focus solely on the damping function and assume $\bar{T}/T_w = 1$ (no variable-property effects), meaning we do not solve the energy equation. Additionally, we assume a specific Reynolds number, $Re_\tau = 750$, and verify that the final form of $f(M_t)$ is not significantly influenced by the choice of $Re_\tau$.

From the predicted velocity profiles, we compute the log-law intercept $C$ and observe how it changes with variations in $f(M_t)$. For the SST model, the log-law intercept $C$ is non-linearly related to $f(M_t)$ through the expression:
\begin{equation}
C - 5.2 = 12 f(M_t)^{1.3}. \label{cfmt}
\end{equation}

It is known from prior work \citep{hasan2023incorporating} that $C$ is a linear function of the friction Mach number $M_\tau$:
\begin{equation}
C - 5.2 = 7.18 M_\tau. \label{clin}
\end{equation}
Combining equations~\eqref{cfmt} and~\eqref{clin}, we obtain:
\begin{equation}
f(M_t)^{1.3} = \frac{7.18}{12} M_\tau. \label{cfmt2}
\end{equation}

Using DNS data from over 30 cases in the literature, we note that there exists a linear relationship between $M_\tau$ and the maximum value of $M_t$, given by $M_t^{max} \approx 3.33 M_\tau$ (not shown). Substituting this relationship into equation~\eqref{cfmt2}, we derive:
\[
f(M_t)^{1.3} \approx \frac{7.18}{12 \times 3.33} M_t^{max},
\]
which simplifies to:
\begin{equation}
f(M_t) \approx 0.27 \left(M_t^{max}\right)^{0.77}. \label{cfmt3}
\end{equation}

The constant $0.27$ in equation~\eqref{cfmt3} is based on the DNS-derived relationship between $M_\tau$ and $M_t^{max}$. However, the $k$-$\omega$ SST model predicts values of $k$, and hence, $M_t^{max}$ that differ from those obtained from DNS. Moreover, $M_t^{max}$ is a single value, whereas we seek a relationship that depends on the local $M_t$, which varies throughout the domain. To address these issues, we retain the functional form of equation~\eqref{cfmt3} but replace $M_t^{max}$ with the local $M_t$. We then adjust the constant $0.27$ until the relationship $C - 5.2 = 7.18 M_\tau$ is accurately reproduced by the SST model. This process yields:
\begin{equation}
f(M_t) \approx 0.39 M_t^{0.77}. \label{cfmt4}
\end{equation}
The same approach can be used to formulate and tune $f(M_t)$ for other turbulence models.

\section{{The proposed corrections for the Spalart-Allmaras (SA) model}}\label{Sec:appSA}

In this section, we report the variable-property and intrinsic compressibility corrections for the SA model \citep{spalart1992one}, and present the results.

Following the same approach as described in section~\ref{Sec:derivInner} for the inner layer 
and in Appendix~\ref{Sec:appouter} for the outer layer, but now applied to the SA model, we obtain \begin{equation}\label{sain}
\begin{split}
\Phi^{\textrm{in}}_{\check\nu} = &\frac{c_{b 2}}{c_{b 3}}\left( \frac{S_y}{\sqrt{\bar \rho}} \frac{d (\bar\rho/\bar\mu) \check{v}}{d y}\right)^2 - \frac{c_{b 2}}{c_{b 3}}\left(\frac{d \check{v}}{\partial y}\right)^2+\\
    & \frac{1}{c_{b 3}} \frac{S_y}{\bar\rho}\frac{d}{dy}\left[\left(\frac{\bar\mu}{\bar\rho}+\check{v}\right)\frac{\bar\rho}{\bar\mu} S_y \frac{d (\bar \rho/\bar\mu)\check{v}}{dy}\right] - \frac{1}{c_{b 3}} \frac{d}{d y}\left[\left(\frac{\bar\mu}{\bar\rho}+\check{v}\right) \frac{d \check{v}}{dy}\right],
\end{split}
\end{equation}
in the inner layer, and
\begin{equation}\label{saout}
\begin{split}
    \Phi^{\textrm{out}}_{\check\nu} = &\frac{c_{b 2}}{c_{b 3}}\left( \frac{1}{\sqrt{\bar \rho}} \frac{d \sqrt{\rho} \check{v}}{d y}\right)^2 - \frac{c_{b 2}}{c_{b 3}}\left(\frac{d \check{v}}{\partial y}\right)^2+\\
    & \frac{1}{c_{b 3}} \frac{1}{\bar\rho}\frac{d}{dy}\left[\left(\frac{\bar\mu}{\bar\rho}+\check{v}\right)\sqrt{\bar\rho}\frac{d \sqrt{\bar\rho}\check{v}}{dy}\right] - \frac{1}{c_{b 3}} \frac{d}{d y}\left[\left(\frac{\bar\mu}{\bar\rho}+\check{v}\right) \frac{d \check{v}}{dy}\right],
\end{split}
\end{equation}
in the outer layer. $\Phi_{\check \nu}$ in these equations represents the source term that needs to be added to a conventional SA model for appropriate accounting of the variable-property effects, and   $c_{b2}$ and $c_{b 3}$ are the model constants. Note that the outer layer corrections are similar to the ones proposed in \cite{catris2000density} and \cite{otero2018turbulence}. Also note that in these derivations we use $\check\nu^+ = \check\nu/(u_\tau \delta_v)$ and $\check\nu^* = \check\nu/(u_\tau^* \delta_v^*)$ in the inner layer, and $\check\nu^\circledplus = \check\nu/(u_\tau \delta)$ and $\check\nu^\circledast = \check\nu/(u_\tau^* \delta)$ in the outer layer.

\begin{figure}
	\centering	\includegraphics[width=\textwidth]{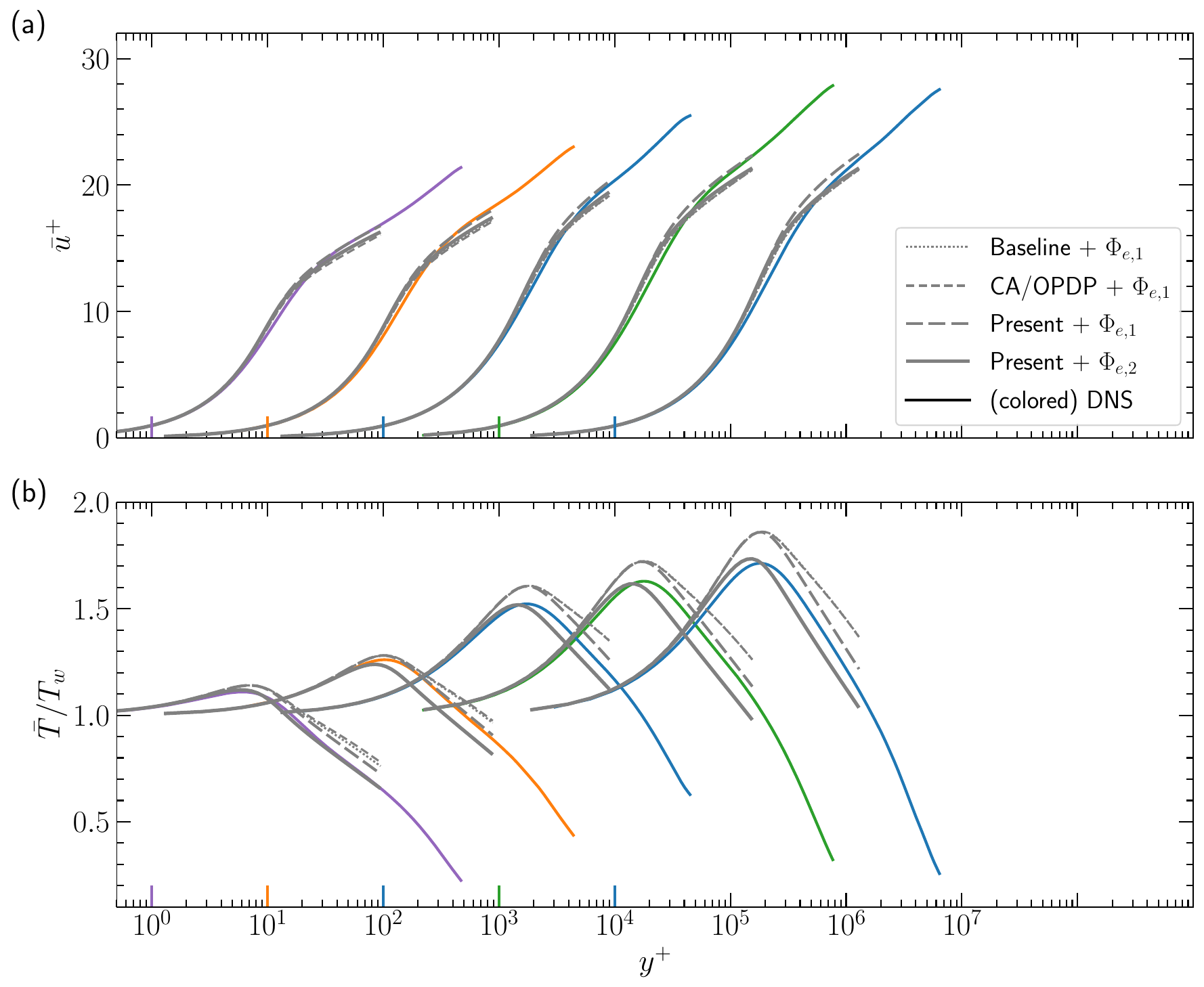}
\caption{{Computed mean velocity (a) and temperature (b) profiles compared to the DNS ({colored} solid lines) for the following boundary layers with increasing wall-cooling: (left to right) $M_\infty=7.87$, $T_w/T_r=0.48$ (A.~Ceci, private communication); $M_\infty=6$, $T_w/T_r=0.35$ \citep{cogo2023assessment}; $M_\infty=5.84$, $T_w/T_r=0.25$ \citep{zhang2018direct}; $M_\infty=10.9$, $T_w/T_r=0.2$ \citep{huang2022direct}; $M_\infty=13.64$, $T_w/T_r=0.18$ \citep{zhang2018direct}. Refer to the legend for line types. `Baseline' stands for the SA model without corrections, `CA/OPDP' stands for the compressibility corrections proposed in \cite{catris2000density, pecnik2017scaling, otero2018turbulence}, `Present' stands for the corrections proposed in this paper (equations~\ref{sain}~and~\ref{Dic_sa}). For clarity, the velocity and temperature profiles for different cases are shifted by one decade along the abscissa. The colored vertical lines on the abscissa signify $y^+=10^0$ for each case, with their colors matching the corresponding cases. 
  }}
 \label{Fig:tblcfdsa}
\end{figure}

To account for intrinsic compressibility effects, similar to the SST model, we propose multiplying the eddy viscosity formulation with a damping function defined as
\begin{equation}\label{Dic_sa}
 (D^{ic})_{\textrm{SA}} = \frac{D(R_t,M_t)}{D(R_t, 0)}, 
\end{equation}
with
\begin{equation}\label{dampsa}
D(R_t,M_t) =  \left[1 - \mathrm{exp}\left(\frac{-R_t}{K + f(M_t)}\right)\right]^2,
\end{equation}
where $K=7.5$, $f(M_t) = 7.3 M_t$ (tuned following a similar procedure as described in Appendix~\ref{Sec:appfmt}), $R_t = \check\nu/(\bar\mu/\bar\rho)$, and $M_t = \sqrt{ (1/0.3) \check \nu d\bar u/dy}/\bar a$ \citep{barone2024data}.

To depict the performance of the proposed corrections compared to the state-of-the-art approaches, we will present the results only for the inner layer of boundary layers. To obtain them we follow the implementation described in section~\ref{impl_bl} but now with $\mu_t$ obtained from the SA model. 
At the wall, we use $\check \nu=0$, whereas at $y/\delta=0.2$, we implement the mixing-length solution for $\check \nu$ as 
\begin{equation}
\begin{array}{ccc}
    \check \nu = {u_\tau^* \kappa y} & \textrm { at } &  y = 0.2\delta. 
\end{array}
\end{equation}

Lastly, to produce results with a more accurate model for the source term ($\Phi_{e,2}$), we define the effective dissipation rate as in equation~\eqref{eps_epsw}, but with $\epsilon_{\mathrm{SA}}$ instead of $\epsilon_{\textrm{sst}}$. To obtain the dissipation rate estimated by the SA model, we adapt the formulation proposed by \cite{rahman2012exploring}, as
\begin{equation}
    \epsilon_{\textrm{SA}} = \frac{ (\mu_t d\bar u/dy)^2/a_1^2}{\bar \mu + \bar \mu_t/a_1^2},
\end{equation}
where $a_1 = 0.3$ is a constant which corresponds to the ratio of the turbulent shear stress and TKE in the log layer \citep{huang1994turbulence}. As desired, $\epsilon_{\textrm{SA}}$ approaches $ \mu_t  (d\bar u/dy)^2$ in the logarithmic region and beyond. 
Refer the provided jupyter-notebook \citep{jupnotebook} for more details on the implementation of the SA model.

Figure~\ref{Fig:tblcfdsa} shows the (a) mean velocity and (b) temperature profiles for the five cases shown in figure~\ref{Fig:tblcfd}. Different line types correspond to different modeling approximations, as discussed earlier in section~\ref{results}. 

Similar to the SST model, the results with the baseline model (without corrections) and those with the CA/OPDP (outer-layer, equation~\ref{saout}) corrections are nearly identical (compare the grey dotted and grey short-dashed lines; in some cases they overlap, making them difficult to distinguish). With these approaches, the mean velocity profiles are quite accurately estimated, whereas the temperature profiles are consistently over-estimated. The accuracy of $\bar u^+$ can be partially attributed to the semi-local consistency of the baseline model itself (see section~\ref{Sec:intro}). However, the high accuracy observed despite not accounting for intrinsic compressibility effects~\eqref{Dic_sa} is likely due to error cancellations resulting from the inaccuracies in the temperature profiles.

With the proposed corrections in equations~\eqref{sain}~and~\eqref{Dic_sa} (long-dashed grey lines), both the temperature and velocity profiles show slight improvement, however, the temperature profiles remain over-estimated, particularly near the peak in strongly cooled boundary layers. The over-estimation of the peak temperature is eliminated, when $\Phi_{e,1}$ is replaced with $\Phi_{e,2}$ (solid grey lines). Despite this improvement, the temperature profiles with $\Phi_{e,2}$ appear to be shifted towards the wall compared to the DNS, resulting in an under-estimation beyond the peak location.

\section*{Acknowledgments}
This work was supported by the European Research Council grant no.~ERC-2019-CoG-864660, Critical. The authors gratefully acknowledge the fruitful discussions with Yuri Egorov (Ansys).
The authors also thank the reviewers for their constructive feedback.

\backsection[Declaration of Interests]{The authors report no conflict of interest.}


\end{document}